\documentclass[journal,twoside,web]{ieeecolor}
\usepackage{tuffc}
\usepackage{cite}
\usepackage{amsmath,amssymb,amsfonts}
\usepackage{algorithm}
\usepackage{url}
\usepackage{algorithmic}
\usepackage{nicefrac}
\usepackage{esint}
\usepackage{algorithmic}
\usepackage{graphicx}
\usepackage{textcomp}
\usepackage{soul}
\usepackage{float}
\usepackage{wrapfig,colortbl}

\usepackage{tikz}
\usepackage{textcomp}
\newcommand\copyrighttext{%
  \footnotesize \textcopyright 2024 IEEE. Personal use of this material is permitted. Permission from IEEE must be obtained for all other uses, in any current or future media, including reprinting/republishing this material for advertising or promotional purposes, creating new collective works, for resale or redistribution to servers or lists, or reuse of any copyrighted component of this work in other works. DOI: 10.1109/TUFFC.2024.3456048.}
\newcommand\copyrightnotice{%
\begin{tikzpicture}[remember picture,overlay]
\node[anchor=south,yshift=10pt] at (current page.south) {\fbox{\parbox{\dimexpr\textwidth-\fboxsep-\fboxrule\relax}{\copyrighttext}}};
\end{tikzpicture}%
}

\definecolor{abstractbg}{rgb}{1,0.969,0.914}
\def\BibTeX{{\rm B\kern-.05em{\sc i\kern-.025em b}\kern-.08em
    T\kern-.1667em\lower.7ex\hbox{E}\kern-.125emX}}
\begin{document}

\title{Focal Volume, Acoustic Radiation Force, and Strain in Two-Transducer Regimes}
\author{Kasra Naftchi-Ardebili, Mike D. Menz, Hossein Salahshoor, Gerald R. Popelka, Stephen A. Baccus, Kim Butts Pauly
\thanks{This work was supported by the NIH R01 Grant EB032743.}
\thanks{K. Naftchi-Ardebili is with the Department of Bioengineering, Stanford University, Stanford, CA 94305 USA. (e-mail: knaftchi@stanford.edu).}
\thanks{M. D. Menz is with the Department of Neurobiology, Stanford University, Stanford, CA 94305 USA. (e-mail: mdmenz@stanford.edu).}
\thanks{H. Salahshoor is with the Departments of Civil and Environmental Engineering, and Mechanical Engineering and Materials Science, Duke University, Durham, NC 27710 USA. (e-mail: hossein.salahshoor@duke.edu).}
\thanks{G. R. Popelka is with the Departments of Otolaryngology and Radiology, Stanford University, Stanford, CA 94305 USA. (e-mail: gpopelka@stanford.edu).}
\thanks{S. A. Baccus is with the Department of Neurobiology, Stanford University, Stanford, CA 94305 USA. (e-mail: baccus@stanford.edu).}
\thanks{K. Butts Pauly is with the Departments of Radiology, Bioengineering, and Electrical Engineering, Stanford University, Stanford, CA 94305 USA. (e-mail: kimbutts@stanford.edu).}
\copyrightnotice
}

\IEEEtitleabstractindextext{%
\fcolorbox{abstractbg}{abstractbg}{%
\begin{minipage}{\textwidth}\rightskip2em\leftskip\rightskip\bigskip
\begin{wrapfigure}[15]{r}{3in}%
\hspace{-3pc}\includegraphics[width=2.9in]{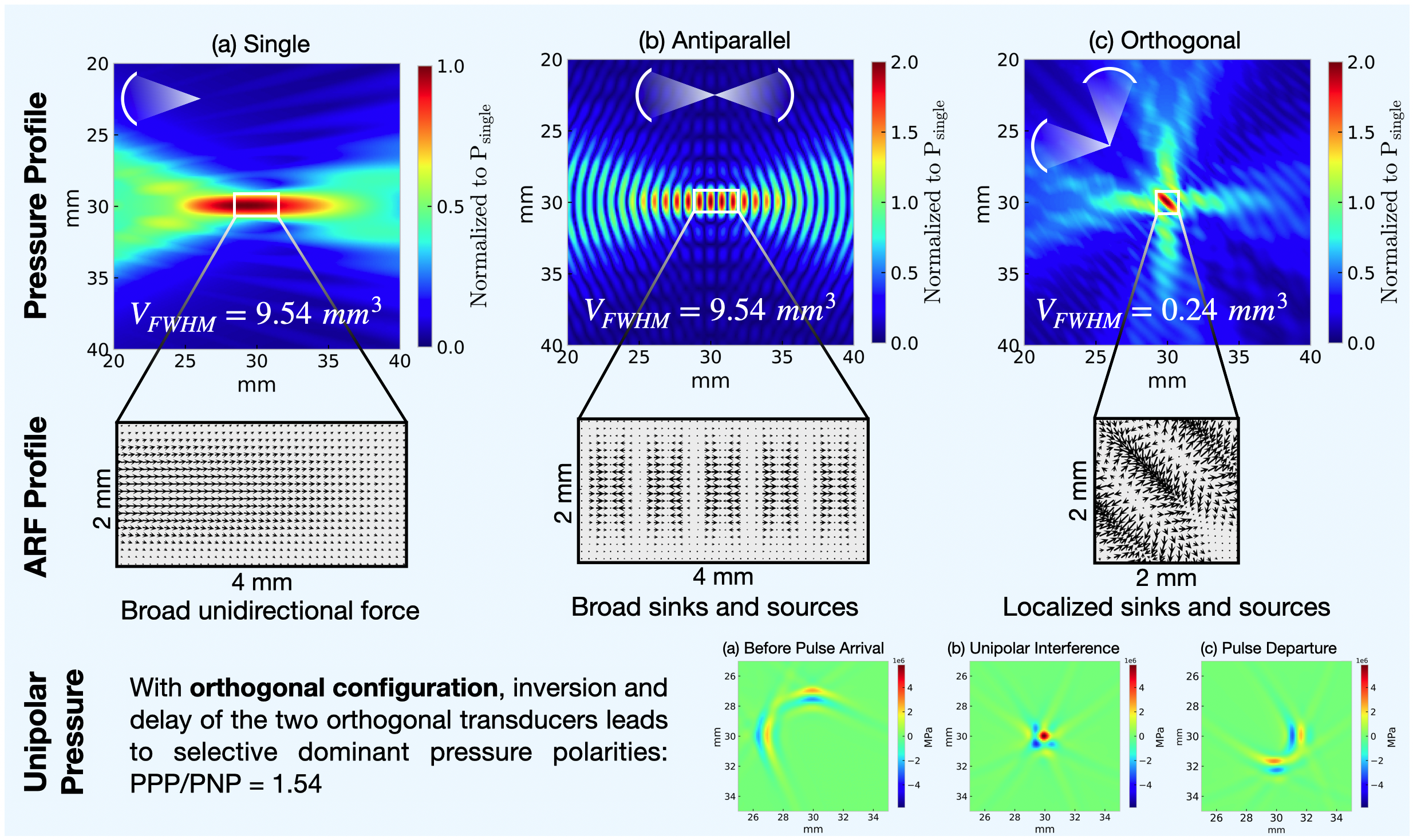}
\end{wrapfigure}%
\begin{abstract}
Transcranial focused ultrasound stimulation (TUS) holds promise for non-invasive neural modulation in treating neurological disorders. Most clinically relevant targets are deep within the brain (near or at its geometric center), surrounded by other sensitive regions that need to be spared clinical intervention. However, in TUS, increasing frequency with the goal of improving spatial resolution reduces the effective penetration depth. We show that by using a pair of 1 MHz, orthogonally arranged transducers we improve the spatial resolution afforded by each of the transducers individually, by nearly 40 fold, achieving a sub-cubic millimeter target volume of $0.24\ mm^3$. We show that orthogonally placed transducers generate highly localized standing waves with Acoustic Radiation Force (ARF) arranged into periodic regions of compression and tension near the target. We further present an extended capability of the orthogonal setup, which is to impart selective pressures--either positive or negative, but not both--on the target. Lastly, we share our preliminary findings that strain can arise from both particle motion and ARF with the former reaching its maximum value at the focus, and the latter remaining  null at the focus and reaching its maximum around the focus. 

As the field is investigating the mechanism of interaction in TUS by way of elucidating the mapping between ultrasound parameters and neural response, orthogonal transducers expand our toolbox by making it possible to conduct these investigations at much finer spatial resolutions, with localized and directed (compression vs. tension) ARF and the capability of applying selective pressures at the target.
\end{abstract}

\bigskip
\end{minipage}}}

\maketitle

\section{Introduction}
\label{sec:introduction}
\IEEEPARstart{S}{ince} the early experiments by Fry et al.\cite{FRY1958IntenseSystem.}, interest in ultrasound as a noninvasive method for high resolution deep brain stimulation has grown substantially. Recent large animal studies with single element or phased array transducers have shown important results and strong evidence toward this goal \cite{Yang2018NeuromodulationDetection, Folloni2019ManipulationStimulation, Kim2021TranscranialSheep, Legon2018NeuromodulationThalamus, Mohammadjavadi2022TranscranialMR-ARFI}. While somewhat large focal spots, on the order of $10\ mm^3$ at full width at half maximum (FWHM) of intensity, achievable with single element transducers are reasonable, and even desirable, for many applications there is still a need in pushing the envelope of resolution. High spatial resolution in focused ultrasound neuromodulation and ablation is essential for precise targeting of specific brain regions, minimizing off-target effects, and enhancing therapeutic outcomes. This precision allows for effective modulation of neural circuits, enabling significant alterations in neural activity and behavior, particularly in deep brain structures like the thalamus, and is crucial for applications such as treating neurological disorders or studying specific brain functions \cite{Zhang2021TranscranialAnimals, Sanguinetti2020TranscranialHumans}. Emphasizing the significance of reaching smaller target volumes, Martin et al. developed a helmet transducer comprising 256 elements, with a FWHM focal volume of $24\ mm^3$ at 555 kHz, approximately 1000 times smaller than what is achievable with small aperture conventional transducers \cite{MartinUltrasoundCircuits}.
There are roughly $50,000$ neurons in any cubic millimeter of cerebral cortex in humans\cite{MacKinnon2022Sensitivity}. Sonicating a $10$-cubic millimeter target with a single element 1 MHz transducer at a HM intensity will engage around $500,000$ neurons. Stimulation of fewer numbers of neurons will characterize neural subpopulation responses to TUS on a finer scale. 

Prior work identified the gain in spatial resolution accompanied by a pair of transducers with overlapping beams in vibro-acoustography \cite{chen2004comparison}, and more specifically an orthogonal pair of transducers in focused ultrasound stimulation  \cite{Kim2018ImprovedTransducers, Yang2020DevelopmentStudy, Kim2021TranscranialResolution}. However, as regards focused ultrasound stimulation, these efforts had only scratched the surface as neither the acoustic radiation force (ARF) nor the standing waves present in the orthogonal arrangement were thoroughly investigated. In fact, historical aversion towards unwanted standing waves that arise from scattering of the ultrasound beams inside the skull compelled authors to deliberately annihilate the standing waves in their orthogonal pair of transducers \cite{Yang2020DevelopmentStudy}. Interestingly, in their 2019 paper, Menz et al. \cite{Menz2019RadiationRetina} showed that stronger neural responses were recorded when salamander retina was subjected to standing ultrasound waves \emph{in vitro}, in contrast to propagating waves \emph{in vivo}. They observed significantly stronger neural activity at transducer-multi electrode array distances that were $0.5$ multiples of the wavelength, where standing waves occur, and validated the important role standing waves may play in eliciting neural responses in the salamander retina \emph{in vitro}. To leverage this interesting finding, Kim et al. \cite{Kim2022PatternedNeuromodulation} proposed the use of two single element transducers \emph{in vivo}, on opposite sides of the skull facing each other in what we call the antiparallel setup, to generate standing waves at the neural target of interest. Although the most intuitive approach in replicating the \emph{in vitro} standing waves observed and validated by Menz et al.\cite{Menz2019RadiationRetina}, a pair of antiparallel transducers failed to improve the spatial resolution afforded by either transducer because the broad axial beam profiles of the two transducers overlapped with one another producing yet another broad axial profile. 

In this work, we show that placement of two single element transducers at a $90^{\circ}$ angle to each other (orthogonal arrangement) achieves both objectives: it increases spatial resolution  by reaching sub-cubic millimeter target volumes ($0.24\ mm^3$ in our setup, roughly $12,000$ neurons in contrast to $500,000$), and simultaneously generates standing waves that were previously reported effective by Menz et al. 

In this work, we consider not only the control of the sonication target volume but also the underlying physical effects that give rise to neural stimulation. There is growing evidence that neurons respond to the mechanical effects of ultrasound, either from the pressure, acoustic radiation force (ARF) effects, strain, or even an ensemble of these \cite{Wright2017UnmyelinatedUltrasound, Kubanek2018UltrasoundSystem, Menz2019RadiationRetina, Blackmore2019UltrasoundSafety, Prieto2013DynamicForce, Kamimura2020UltrasoundAssessment}. In this report, we first describe the mathematical models of ARF at three different tiers of complexity and report their accuracy for a single transducer. We do this to show the inconsistencies of the two simpler models, because various forms of ARF formulae are assumed in the literature and sometimes the underlying assumptions are not explicitly stated \cite{Kim2022PatternedNeuromodulation, Palmeri2011AcousticMethods, Hsu2005AcousticProbe, Mcaleavey2009ValidationHydrogels, Haar2010UltrasoundSafety, Nightingale2003Shear-waveResults, Nightingale2006AnalysisForce, Doherty2013AcousticUltrasound, Zhang2018FromForces, Sarvazyan2021AcousticApplications}. We believe it is critical to discuss these various models in detail because adoption of these simplifying assumptions without regards to whether the assumed conditions are met, gives rise to erroneous ARF calculations, both in terms of magnitude and direction. Next, we show the ARF for the three setups of single transducer, antiparallel transducers, and orthogonal transducers without the simplifying assumptions that fail in the limit of curved (focused) wavefronts. By computing the ARF in the orthogonal setup, we demonstrate interesting properties of the force field, notably the locations of the force sinks and sources. Although these periodic sinks and sources are also present in the antiparallel setup, they are not localized and span the axial profile of the two transducers. Because neural populations may respond selectively to tension (at ARF source) or compression (at ARF sink), this type of modeling may then become crucial for controlled transcranial neuromodulation. Insofar as neuromodulation is concerned, the control of pressure at the focus typically has been limited to modulating the pressure amplitudes and pulsing schemes. With these conventional modulations spatial and temporal periodicity of the wave remain intact; that is to say, every rarefaction is followed by a countervailing compression, within the limits of a linear regime. We currently don’t know whether in neurological diseases where asynchronous  or abnormal firing of the neurons is the underlying cause, being able to selectively subject the neurons to rarefaction only or compression only may prove optimal. Selective application of pressure at non-destructive levels--in contrast to destructive levels such as shock waves--has never been attempted before. We introduce the unipolar pressure method that allows us to select, in advance, the polarity of the dominant pressure (either positive pressure only or negative pressure only) we would like to deliver to the target of interest. Such a technique, possible only with orthogonal transducers of high bandwidth (single or tricycle pulses), would allow us to study the pressure preference of neurons in different regions of the brain and explore the existence of pressure-specific responses in neural populations.

Lastly, we present our preliminary findings on strain and show the different fields that arise from strain due to particle motion versus strain due to ARF. These strains are related as they occur simultaneously, but one may dominate the other depending on the choice of frequency. 

Simultaneous and accurate knowledge of pressure, ARF, and strain for the region being sonicated is essential in elucidating the ultrasound-neuron mechanism of interaction, and is indispensable if the ultimate goal is to develop a valid mapping between TUS parameters and neural activity. It is the aim of this study to shed light on these unknowns.

\begin{table*}[!t]
\arrayrulecolor{subsectioncolor}
\setlength{\arrayrulewidth}{1pt}
{\sffamily\bfseries\begin{tabular}{lp{6.75in}}\hline
\rowcolor{abstractbg}\multicolumn{2}{l}{\color{subsectioncolor}{\itshape
Highlights}{\Huge\strut}}\\
\rowcolor{abstractbg}$\bullet$ & The use of orthogonally arranged transducers in TUS achieves enhanced spatial resolutions and introduces the capability of applying localized acoustic radiation force and selective pressures for targeted neural modulation.\\
\rowcolor{abstractbg}$\bullet${\large\strut} & Orthogonal configuration enhances spatial resolution by 40x and enables localized tension and compression through ARF. It further allows selective pressures with peak positive to peak negative pressure ratio of $1.54$ for precise targeting and control in neuromodulation.\\
\rowcolor{abstractbg}$\bullet${\large\strut} & 
These findings pave the way for more precise and effective treatments for neurological disorders through high resolution targeting and modulation capabilities, using only two orthogonally-placed transducers.\\[2em]\hline
\end{tabular}}
\setlength{\arrayrulewidth}{0.4pt}
\arrayrulecolor{black}
\end{table*}

\section{Methods}
\subsection{Simulation Setup}
All ultrasound simulations were performed in k-Wave \cite{Treeby2010K-Wave:Fields}, a widely used MATLAB \cite{TheMathWorksInc2019MATLABR2019b} package for the time-domain simulation of the acoustic wave fields. Our aim was to show the possibility of reaching sub-cubic millimeters in FWHM target volume using $1$ MHz transducers. With $128$ GB of system memory available to us, we were unable to run realistic transcranial simulations at such high frequencies at depths greater than a couple of centimeters. Therefore, to be able to simulate the $1$ MHz scenarios at $0.1\ mm$ resolution, we limited ourselves to single-element transducers with $30\ mm$ aperture diameter and $30\ mm$ focal length. However, to establish that even in transcranial settings the gain in spatial resolution with orthogonal configuration is significant, we performed one simulation with $84\text{-}mm$ wide, $420$-element phased array transducers at $250$ kHz and $0.2\ mm$ resolution. For this scenario, phase aberration correction was done using time reversal \cite{Fink1994PhaseMirrors}.

Specific simulation parameters are presented in Table \ref{tab:naftc.t1}.

\begin{table}[H]
\caption{k-Wave Parameters for Different Simulation Setups}
\label{table}
\setlength{\tabcolsep}{3pt}
\resizebox{\linewidth}{!}{%
\begin{tabular}{*{6}{p{\dimexpr(\linewidth-12\tabcolsep)/6}}}
\hline\hline
Simulation\par Setup & 
Pulse \par Form & 
Pressure\par Amplitude & 
Frequency & 
Resolution& Acoustic \par Velocity\\
\hline
Free Water\par FWHM\par Resolution\par & 
Continuous\par Wave & 
1 MPa & 
1 MHz & 
0.1 mm & 
1500 m/s \\ 
Transcranial\par FWHM\par Resolution\par & Continuous\par Wave & 
0.5 MPa & 
250 kHz & 
0.2 mm & 
1500 -\par 3175 m/s \\ 
ARF & 
Continuous\par Wave & 
1 MPa & 
1 MHz & 
0.1 mm & 
1500 m/s \\ 
Unipolar\par Pressure\par & 3-Cycle,\par Gaussian & 1 MPa & 1 MHz & 0.1 mm & 1500 m/s\\
PM Strain\par & 20-Cycle,\par Gaussian & 5 MPa & 250 kHz & 0.2 mm & 1500 m/s\\
ARF Strain & Step Force & 5 MPa & 250 kHz & 0.2 mm & 20 m/s\\
\hline\hline
\multicolumn{6}{p{\dimexpr\linewidth-2\tabcolsep}}{Transcranial simulation required a larger field to capture a transverse slice of human skull CT in full. To reduce the computation load, we ran this simulation at $0.2\ mm$ resolution and at a lower frequency of $250$ kHz. We chose $5$ MPa for strain simulations to be able to observe a sharp and clear tissue displacement. Because ARF strain simulations were run for $5\ ms$, we reduced the grid resolution to $0.2\ mm$ and the acoustic velocity to $20\ \nicefrac{m}{s}$, to reduce the computational load of these simulations along the temporal dimension. These considerations would not affect the normalized ARF strain maps.}\\
\end{tabular}}
\label{tab:naftc.t1}\\
\end{table}

Whereas ARF and unipolar pressure simulations were with $1$ MHz transducers, for strain simulations we set the center frequency to $0.25$ MHz to reduce the computational load along the temporal axis. For the same reason, following the approach presented by Prieur et al. \cite{Prieur2016Simulationk-Wave}, compressional acoustic velocity for ARF strain simulations was set to $20\ \nicefrac{m}{s}$ (instead of $1500\ \nicefrac{m}{s}$ in ARF and unipolar pressure simulations), so that it was computationally feasible to simulate the slower-timescale tissue displacements ($5\ ms$ in our setup). In k-Wave, bulk and shear moduli are computed from compressional and shear wave velocities, respectively. Therefore, reducing the compressional velocity has the implicit effect of reducing the bulk modulus of the tissue, which in turn makes the medium less resistant to compression. However, given we are only reporting normalized values at steady state, subsequent comparisons between strain due to ARF and strain due to particle motion are still valid.
Without these modifications, simulation memory requirements exceeded the 128 GB of memory we had available to us. Given our simulation conditions, namely the relatively low pressure amplitude ($1$ MPa) in ARF and unipolar pressure scenarios, and a homogeneous medium, nonlinear effects were minimal. Nonetheless, for all strain simulations under 5 MPa, we incorporated nonlinearity into our simulations. It should be emphasized that simulations in k-Wave were performed on a collocated grid, because pressure and particle velocities were simultaneously required to compute the ARF \cite{Prieur2016Simulationk-Wave}. As such, particle velocities must be obtained via $\tt{u\_non\_staggered}$ in k-Wave, otherwise the default option simulates the staggered velocities, resulting in erroneous ARF calculations.

\subsection{Transducer Arrangements}
Three transducer arrangements were studied: a single transducer, two transducers in an antiparallel configuration, and two transducers at orthogonal angles. In the antiparallel setup, two identical transducers were placed on-axis, with their focal points overlapping. In the orthogonal setup, two transducers were placed at $90^{\circ}$ with respect to one another with their focal points overlapping. 

In computing the target volumes, each of three arrangements were compared to the other two, both in free water and transcranial settings. In investigating the effect of the simplifying assumptions on ARF, only the case of a single transducer was studied. In comparing the ARF field in single, antiparallel, and orthogonal configurations, only the complete form of ARF, equation (\ref{eq:ARF_complete}), was employed. Unipolar pressures were presented for orthogonal transducers only, as they are not feasible with other arrangements. And lastly, we report our preliminary results on strain for all three configurations.

\subsection{Spatial Resolution}
To compute the spatial resolution we assumed an ellipsoidal target. Axis lengths for the ellipsoid were obtained through the FWHM of the intensity at $-3$ dB, which were then used to compute the volume of the ellipsoid with the following equation:

\begin{equation}\label{eq:volume}
    V = \frac{4}{3}\pi \bigg(\frac{a}{2}\bigg)\bigg(\frac{b}{2}\bigg)^2,
\end{equation}

\noindent where $a$ and $b$ were the major and minor axes of the ellipsoid, respectively. Although the simulations were conducted in 2D where we had only one major and one minor axis for an ellipse, the symmetry of the problem dictated that addition of a $3^{rd}$ axis would introduce yet another minor axis. This allowed us to estimate the 3D target volume from a 2D simulation simply by introducing an additional $\big(\frac{b}{2}\big)$ in equation (\ref{eq:volume}), hence the power of $2$.

\subsection{Acoustic Radiation Force}
The impulse of a force applied to a control volume results in change of momentum in the control volume:

\begin{align}\label{eq:impulse}
    \frac{d\textbf{P}_V}{dt} &= \textbf{F},
\end{align}

\noindent whereby force vector \textbf{F} could be directly computed from the rate of change of momentum vector $\textbf{P}_V$ in the control volume of interest. Note that vector quantities are in bold font. The left-hand side (LHS) in equation (\ref{eq:impulse}) could be written out explicitly as:

\begin{equation}\label{eq:momentum_flux}
    \frac{d\textbf{P}_V}{dt} = \frac{d\textbf{P}}{dt} + \frac{d\textbf{P}_{\text{out}}}{dt} - \frac{d\textbf{P}_{\text{in}}}{dt}.
\end{equation}

The right-hand side (RHS) requires some explanation \cite{MarkDrela2005FluidMechanics, JohnD.Anderson2001FundamentalsAerodynamics}: 

\begin{itemize}
    \item $\frac{d\textbf{P}}{dt}$: rate of change of instantaneous momentum inside the control volume, where $\textbf{P}(t)=\iiint\rho \textbf{v}dV$. This force component is equivalent to the ARF. 
    \item $\frac{d\textbf{P}_{\text{out}}}{dt}$: rate of momentum leaving the control volume due to mass flow.
    \item $\frac{d\textbf{P}_{\text{in}}}{dt}$: rate of momentum entering the control volume due to mass flow.
\end{itemize}

Substituting \textbf{F} for the LHS in equation (\ref{eq:momentum_flux}), and recognizing that $\frac{d\textbf{P}_{\text{out}}}{dt} - \frac{d\textbf{P}_{\text{in}}}{dt}$ is momentum flux through a closed area $A$, that is $\varoiint \rho(\textbf{v}\cdot\hat{\textbf{n}})\textbf{v}dA$, we can rewrite equation (\ref{eq:momentum_flux}) as:

\begin{equation}\label{eq:momentum_flux_integrals}
    \textbf{F} = \frac{d}{dt}\iiint\rho \textbf{v}dV + \varoiint \rho(\textbf{v}\cdot\hat{\textbf{n}})\textbf{v}dA.
\end{equation}

We can expand the LHS into body forces, notably  the gravity, and surface forces, pressure and viscosity, to write out the integral momentum equation \cite{MarkDrela2005FluidMechanics, JohnD.Anderson2001FundamentalsAerodynamics}:

\begin{equation}\label{eq:integral_momentum}
    \begin{split}
        \iiint \rho\textbf{g}dV - \varoiint p\hat{\textbf{n}}dA + \textbf{F}_{\text{viscous}} &= \\
        \frac{d}{dt}\iiint\rho \textbf{v}dV + \varoiint \rho(\textbf{v}\cdot\hat{\textbf{n}})\textbf{v}dA,
    \end{split}
\end{equation}

\noindent where the LHS terms are gravitational force, pressure force, and force due to viscosity, respectively. Note that $\textbf{g}$ is the acceleration due to gravity, $\rho$ is density, $p$ is pressure, and $\hat{\textbf{n}}$ is the normal vector to the surface $A$. Since the timescale over which particle collisions take place is very short, we can ignore the gravitational force. Moreover, because the bulk modulus of biological tissue is much greater than the shear modulus, and that normal stresses are dominant compared to shear stresses, it is reasonable to drop the viscous force \cite{Prieur2017ModelingElastography}. Momentum flux and force due to pressure are in the form of surface integrals, whereas ARF is a volume integral. To be able to assume an infinitesimal volume for both the LHS and the RHS, we need to first rewrite the momentum flux and force due to pressure in terms of volume integrals. In order to convert the momentum flux and pressure force to volume integrals, we invoked the divergence theorem and gradient theorem, respectively. That allowed us to write everything in terms of volume integrals, and drop the integrals with the intention of studying an infinitesimal $\delta V$ volume. Complete derivation is included in the Appendix, where we show how we arrive at the following equation for the $i^{th}$ component of ARF:

\begin{equation}\label{eq:ARF_component}
    ARF(t)_i = -\frac{\partial}{\partial x_k}\big(p\delta_{ik} + \rho v_iv_k \big),
\end{equation}

\noindent where $\delta$ denotes the Kronecker delta, $p$ is pressure, $\rho$ is unperturbed density (constant), $v$ is particle velocity, and $x_k$ is the $k^{th}$ dimension in space. Due to the periodicity of the waves, integrating equation (\ref{eq:ARF_component}) over one complete cycle is sufficient to compute the net ARF \cite{Prieur2017ModelingElastography}:

\begin{equation}\label{eq:ARF_net}
    ARF_i = -\frac{\partial}{\partial x_k}\big(\langle p \rangle\delta_{ik} + \langle \rho v_iv_k \rangle\big),
\end{equation}

\noindent with $\langle \cdot \rangle = \frac{1}{\pi}\int_{t'}^{t' + \pi}(\cdot)dt$, where $\pi$ denotes the period. Note that repeated indices in $k$ follow Einstein summation. Invoking the first law of thermodynamics and utilizing Taylor's expansion make it possible to expand the $\langle p \rangle$ term to second order approximation and write the general form of $ARF_i$ as follows \cite{Prieur2017ModelingElastography, Lee1993AcousticPressure}:

\begin{equation}\label{eq:ARF_net_explicit}
    ARF_i = -\frac{\partial}{\partial x_k}\bigg[ \bigg( \frac{1}{2\rho c^2}\langle p^2 \rangle\ - \frac{1}{2}\rho\langle |v|^2 \rangle \bigg) \delta_{ik} + \langle \rho v_iv_k \rangle\bigg],
\end{equation}

\noindent with $c$ and $v$ denoting the sound and particle velocities. The two terms inside the parentheses refer to the mean Eulerian excess pressure \cite{Lee1993AcousticPressure}, while the last term is the Reynolds stress tensor. We will investigate the complete form of equation (\ref{eq:ARF_net_explicit}), as well as its reduced forms under two tiers of quasi-planar wave and purely-planar wave simplifying assumptions.

\subsubsection{Complete Form}
The simulations in this work were performed in 2D. Substituting $x$ and $y$ for indices in equation (\ref{eq:ARF_net_explicit}) to emphasize that simulations were in Cartesian coordinates, we arrive at the $x$ and $y$ components of ARF:

\begin{equation}\label{eq:ARF_complete}
    \begin{pmatrix}
    F_x \\
    F_y
    \end{pmatrix}
    =
    -
    \begin{pmatrix}
    \begin{split}
        \frac{\partial}{\partial x} \bigg( \frac{1}{2\rho c^2}\langle p^2 \rangle\ - \frac{1}{2}\rho\langle v_x^2 + v_y^2 \rangle \bigg) &+\\
        \rho \big\langle  2v_x\frac{\partial v_x}{\partial x} + v_x\frac{\partial v_y}{\partial y} + v_y\frac{\partial v_x}{\partial y}\big\rangle \\
        \\
        \frac{\partial}{\partial y} \bigg( \frac{1}{2\rho c^2}\langle p^2 \rangle\ - \frac{1}{2}\rho\langle v_x^2 + v_y^2 \rangle \bigg) &+ \\
        \rho \big\langle  2v_y\frac{\partial v_y}{\partial y} + v_y\frac{\partial v_x}{\partial x} + v_x\frac{\partial v_y}{\partial x}\big\rangle
    \end{split}
    \end{pmatrix}.
\end{equation}

\subsubsection{Quasi-Planar Wave Assumption}
This tier of simplifying assumptions considers $p=\rho c v$, which holds for planar waves. Substituting $\rho cv$ for $p$ in the $\langle p^2\rangle$ term in equation (\ref{eq:ARF_complete}) annihilates the mean Eulerian excess pressure. The resulting ARF formula becomes: 

\begin{equation}\label{eq:ARF_quasi}
    \begin{pmatrix}
    F_x \\
    F_y
    \end{pmatrix}^{\text{QP}}
    = \\
    -
    \begin{pmatrix}
    \rho \langle  2v_x\frac{\partial v_x}{\partial x} + v_x\frac{\partial v_y}{\partial y} + v_y\frac{\partial v_x}{\partial y}\rangle \\
     \rho \langle  2v_y\frac{\partial v_y}{\partial y} + v_y\frac{\partial v_x}{\partial x} + v_x\frac{\partial v_y}{\partial x}\rangle
    \end{pmatrix}.
\end{equation}

\begin{figure*}[ht]
\centering
\includegraphics[width=\linewidth]{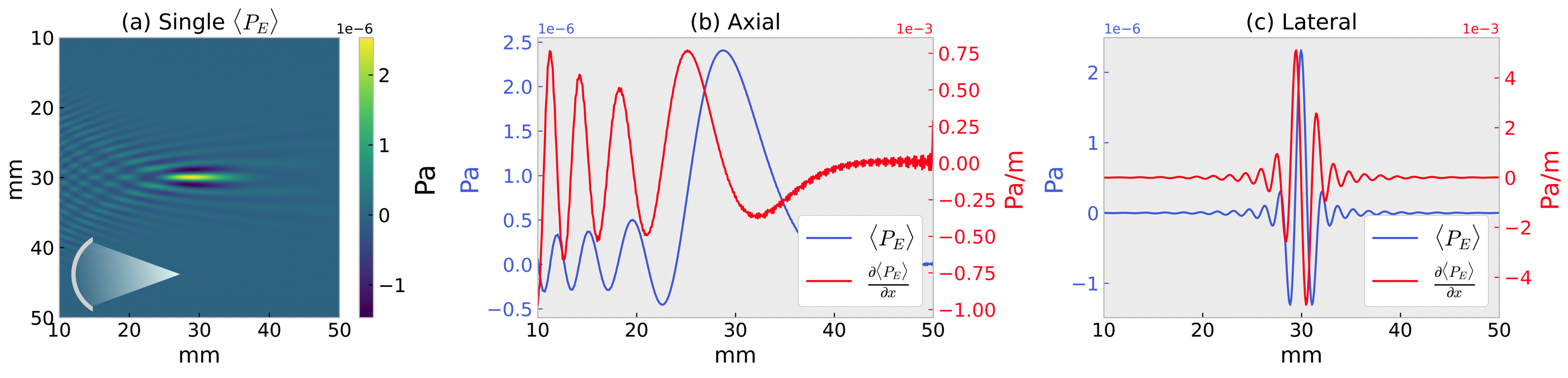}
\caption{Terms that would be ignored in the quasi-planar wave assumption. \textbf{a}) The mean Eulerian excess pressure. Transducer orientation is indicated in the lower left hand corner of the figure. \textbf{b}) Axial beam profiles of the mean Eulerian excess pressure as well as its gradient. \textbf{c}) Lateral beam profiles of the mean Eulerian excess pressure as well as its gradient. Note that even though the mean Eulerian excess pressure itself may seem quite small and negligible, its gradients are about three orders of magnitude larger than the mean Eulerian excess pressure itself. Therefore, neglecting these terms results in erroneous ARF computations.}
\label{fig:naftc1}
\end{figure*}

While considerably simplifying equation (\ref{eq:ARF_complete}), the quasi-planar wave assumption drops consequential terms. Fig. \ref{fig:naftc1} shows these neglected terms: the mean Eulerian excess pressure, along with its axial and lateral gradients. Although the mean Eulerian excess pressure itself could be deemed negligible compared to the pressure values at the focus, its gradient is orders of magnitude larger and should not be hastily dropped. The quasi-planar wave assumption could be applied in scenarios where the wavefront curvature is minimal, such as in the far-field region.

\subsubsection{Purely-Planar Wave Assumption}
The simplest form of ARF assumes propagation of purely planar waves, where in addition to substituting $\rho c v$ for the pressure term, any lateral wave propagation is disregarded. As such, in our simulations where a single transducer is sonicating in $x$, any $y$ component of velocity will be ignored. The ARF formula under these assumptions becomes: 

\begin{equation}\label{eq:ARF_planar}
    ARF^{\text{PP}} = 
    F_{x}^{\text{PP}}
    = -
    \rho \bigg\langle  2v_x\frac{\partial v_x}{\partial x} \bigg\rangle.
\end{equation}

Equations (\ref{eq:ARF_complete}), (\ref{eq:ARF_quasi}), and (\ref{eq:ARF_planar}) were used to compute the ARF for a single transducer to investigate the effect of these simplifying assumptions on validity of ARF. To compute the valid ARF in the single, antiparallel, and orthogonal configurations, however, only equation (\ref{eq:ARF_complete}) was used.

\subsection{Unipolar Pressures}
\begin{figure}[ht]
\centering
\includegraphics[width=\linewidth]{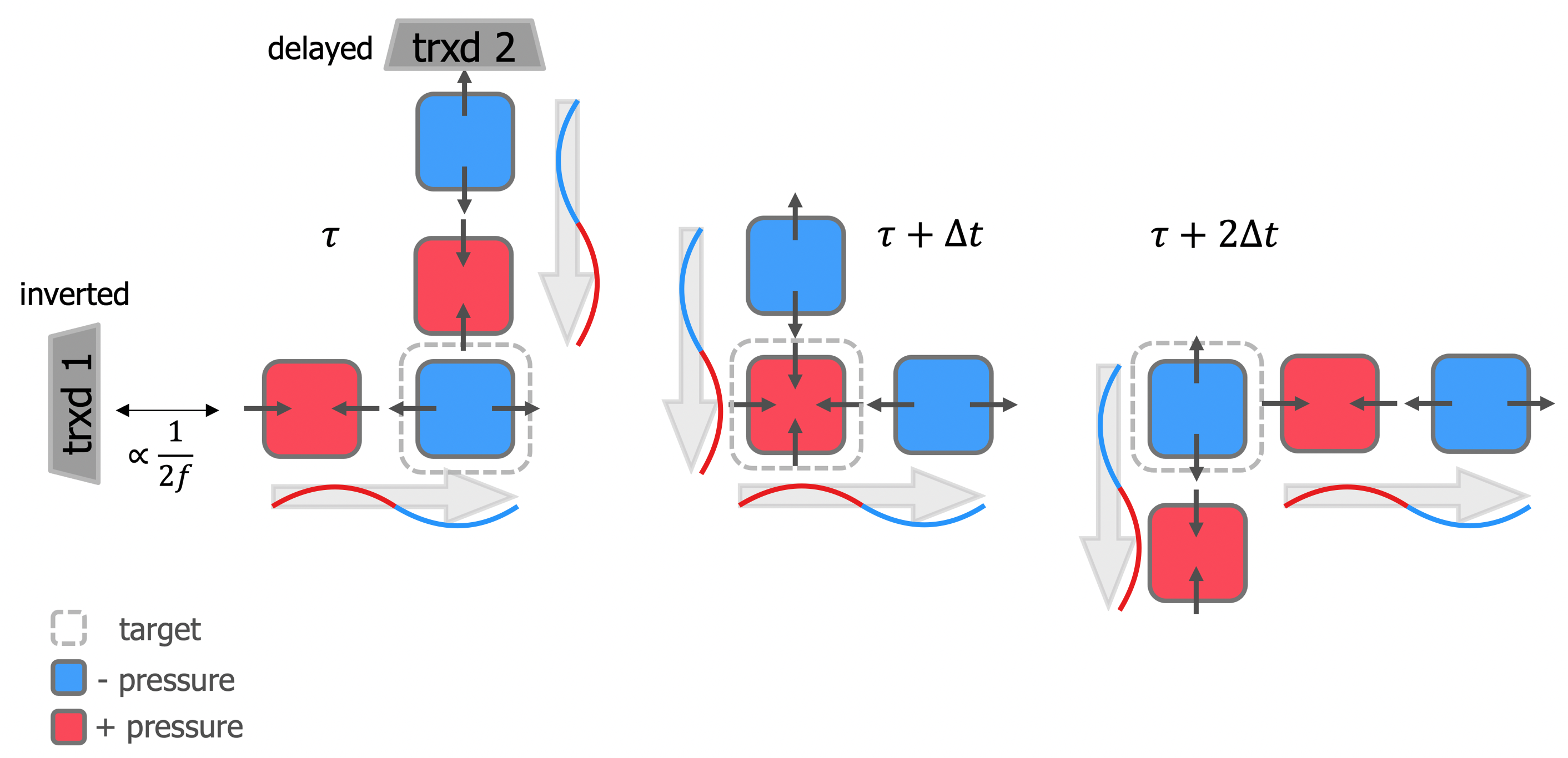}
\caption{Unipolar pressure with orthogonal transducers. Two single-cycle pulses are generated by both transducers. The pulse from transducer 1 is inverted relative to the pulse from transducer 2, while the pulse from transducer 2 is delayed by half the period. At time $\tau + \Delta t$ the two positive poles overlap, resulting in a peak positive to peak negative ratio of 2/1.}
\label{fig:naftc2}
\end{figure}

The approach in imparting selective pressures on the target is best explained with simple schematics, presented in Fig. \ref{fig:naftc2}. In the hypothetical limit of single-cycle pulses, inversion of the pulse from one transducer and delaying of the pulse from the other transducer by half the period will result in either their compression or their rarefaction poles to overlap, but not both. In Fig. \ref{fig:naftc2} the pulse from transducer 1 (trxd 1) is inverted with respect to the pulse from transducer 2 (trxd 2), while the pulse from transducer 2 is delayed by half the period. At time $\tau$, the negative pole of the first transducer arrives at the target in singular form, without overlapping with any pressure fronts from the second transducer. After some $\Delta t$ timesteps, the pressure poles from both transducers overlap at the target, and $2\Delta t$ timesteps later, the two pulses move past the target. This schematic would result in a peak positive to peak negative pressure ratio of \nicefrac{2}{1}, effectively imparting a dominant unipolar positive pressure on the target. 

In an actual simulation, given the curvature of the wave fronts and the window applied to the pulses, implementation of unipolar pressures with simple half-period delay is not feasible. In realistic scenarios an iterative approach, such as the one used in this work (Algorithm \ref{alg:unipolar}), needs to be implemented. The specific transducers we used generated Gaussian-windowed tricycle pulses at 1 MHz center frequency, with their bandwidths spanning from 742 kHz to 1.29 MHz, which corresponds to a fractional bandwidth of approximately 55\%. Optimal delay is one that maximizes the peak positive to peak negative pressure ratio (\nicefrac{PPP}{PNP}) at the target.

\begin{algorithm}
\caption{Calculate optimal delay $n^*$ that maximizes $\frac{PPP}{PNP}$}
\label{alg:unipolar}
\begin{algorithmic} 
\STATE $n$: delay in simulation timesteps
\STATE $n^*$: optimal delay in simulation timesteps
\STATE $\pi$: period in simulation timesteps
\REQUIRE $ 0 \le n \le \pi$
\STATE $p_1$: pressure field for the first transducer, tensor of $t\times x\times y$
\STATE $p_2$: inverted pressure field for the second transducer, tensor of $t\times x\times y$
\STATE $t_i$: initial time (shortly before reaching the target)
\STATE $t_f$: final time (shortly after reaching the target)
\STATE $x^*$: $x$ component of the target 
\STATE $y^*$: $y$ component of the target 
\STATE $ppp\_pnp \gets$ empty list
\FOR{$0 \le i \le int(n) + 1$}
    \STATE $max\_overlap \gets \max(p_1[t_i - i:t_f - i, x^*, y^*] + p_2[t_i:t_f, x^*, y^*])$    
    \STATE $min\_overlap \gets abs(\min(p_1[t_i - i:t_f - i, x^*, y^*] + p_2[t_i:t_f, x^*, y^*]))$        
    \STATE add $max\_overlap / min\_overlap$ to end of $ppp\_pnp$ list
\ENDFOR

\STATE $n^*=\mathrm{argmax}\{ppp\_pnp\}$ 
\end{algorithmic}
\end{algorithm}

\subsection{Strain}
There are generally two routes to computing strain: dynamics and kinematics. To solve for strain through dynamics requires in-depth knowledge of the material properties of the medium, such as knowledge of the accurate relationship between strain and stress. Specifically, ARF could be related to stress via \cite{Prieur2016Simulationk-Wave}:

\begin{equation}\label{eq:arf_stress}
    ARF_i = -\frac{\partial \sigma_{ik}}{\partial_{x_k}},
\end{equation}

\noindent with $\sigma_{ij}$ denoting the stress tensor. In Newtonian fluids such as water, shear stress $\tau$ is linearly related to the shear strain rate \cite{Batchelor2000AnDynamics}: 

\begin{equation}\label{eq:stress_strain_rate}
    \tau = \mu \frac{\partial \gamma}{\partial t},
\end{equation}

\noindent where $\mu$ is the dynamic viscosity term and $\gamma$ is the shear strain. Perhaps putting together equations (\ref{eq:arf_stress}) and (\ref{eq:stress_strain_rate}) would allow computing the ARF by taking the gradient of the strain rate, but that formalism will likely not lend itself to a straightforward calculation of strain itself from ARF. Conversely, through kinematics we can readily solve for strain as long as we have knowledge of displacements along each dimension. In our simulations, we obtained displacement from velocity as follows: 

\begin{equation}\label{eq:displacement}
    u_i[t+1] = u_i[t-1] + 2\Delta t(v_i[t]),
\end{equation}

\noindent where $u_i[t]$ and $v_i[t]$ are position and velocity in the $i^{th}$ direction at time $t$, and $\Delta t$ is the temporal resolution. Strain, in that case, is simply obtained from the Green-Lagrange strain tensor:

\begin{equation}\label{eq:strain}
    \mathcal{E}_{ij} = \frac{1}{2}\bigg( \frac{\partial u_i}{\partial {x_j}} + \frac{\partial u_j}{\partial {x_i}} + \frac{\partial u_k}{\partial x_i}\frac{\partial u_k}{\partial x_j}\bigg),
\end{equation}

\noindent where $\mathcal{E}_{ij}$ represents normal strain when $i=j$ and shear strain when $i\ne j$. Note that the repeated indices are summed over $i$ and $j$. In our simulations, the linear terms were on the order of $10^{-8}$, and the nonlinear terms were on the order of $10^{-16}$. For such small deformations, the nonlinear term is negligible and can safely be ignored\cite{Treeby2014ModellingToolbox}. Nonetheless, we maintained the nonlinear term in our approach and used equation (\ref{eq:strain}) in its entirety.

We have made a clear distinction between strain due to periodic particle motion displacement, which we denote $\mathcal{E}^{\text{PM}}$, and strain due to acoustic radiation force-induced displacement, which we denote $\mathcal{E}^{\text{ARF}}$. The fundamental difference between the two is in how we define displacement, $u$. Under particle motion, the displacement $u^{\text{PM}}$ is defined as the amplitude of the periodic particle oscillations. Under ARF, the displacement $u^{\text{ARF}}$ is defined as the magnitude of the time-averaged bulk tissue displacement. In both cases we used equation (\ref{eq:displacement}) to compute the relevant displacements before applying equation (\ref{eq:strain}). k-Wave readily simulates pressure and particle velocity. Therefore, $\mathcal{E}^{\text{PM}}$ was easily computed by first calling \texttt{kspaceFirstOrder2D}, and then applying equations (\ref{eq:displacement}) and (\ref{eq:strain}). To obtain $\mathcal{E}^{\text{ARF}}$, we need to compute bulk tissue displacements, but k-Wave does not directly simulate these. One indirect way to do this, as show in \cite{Prieur2016Simulationk-Wave}, is to first run \texttt{kspaceFirstOrder2D}, use the pressure and particle velocities to compute the stress, apply that stress as a constant on transducer surface in the \texttt{pstdElastic2D} function, and then follow through with equations (\ref{eq:displacement}) and (\ref{eq:strain}) to obtain $\mathcal{E}^{\text{ARF}}$. Given this indirect approach, we refrain from reporting exact numerical values for $\mathcal{E}^{\text{ARF}}$ and $\mathcal{E}^{\text{PM}}$, but instead normalize each of them independently. It should be emphasized that when we refer to shear strain, we are referring to application of equation (\ref{eq:strain}) to the longitudinal waves under the $i\ne j$ condition, and not application of equation (\ref{eq:strain}) to shear waves themselves. A comparison between strain due to longitudinal and strain due to transverse (shear) waves had previously been reported \cite{Carstensen2016BiologicalDescriptors}. Our focus in this study was the distinction between particle motion strain and ARF strain, both in longitudinal waves. We did not study shear (transverse) waves. 

\subsection{Experimental Setup}
To validate the ARF field in orthogonal configuration, we sonicated glass beads suspended in water, so that it was possible to observe the ARF field distribution with the naked eye. The setup for the glass bead experiment, shown in Fig. \ref{fig:naftc7}(c), consisted of two 1 MHz transducers (Model A303S, 0.5 inch diameter, 0.80 inch focal length, from Olympus) with custom made plastic coupling cones that had a 45-degree bevel. Coupling cones were filled with degassed water. Plastic wrap held down by an o-ring kept the water in place. The transducers were mounted on separate micromanipulators.  Each transducer was mounted on a $x,y,z$ micromanipulator which in turn was mounted on a plate such that two angles could be adjusted.  A dish with a glass bottom was filled with water and the coupling cones were immersed in it. The position and angle of the transducers were manipulated (with no glass beads present) so that not only were their foci overlapping on the glass plate, but also a wave coming from one transducer reflected off the glass bottom and arrived at the second transducer. This was accomplished by using one transducer to send a signal and the other transducer as a receiver to optimize alignment. 50-micron glass beads were placed in a single layer in the bottom of the dish.  A $100\ ms$ continuous wave stimulus was presented once a second. The resulting glass bead movement was captured by a CCD camera mounted from below the dish.

\section{Results}
\subsection{Spatial Resolution}
\begin{figure*}[ht]
\centering
\includegraphics[width=\textwidth]{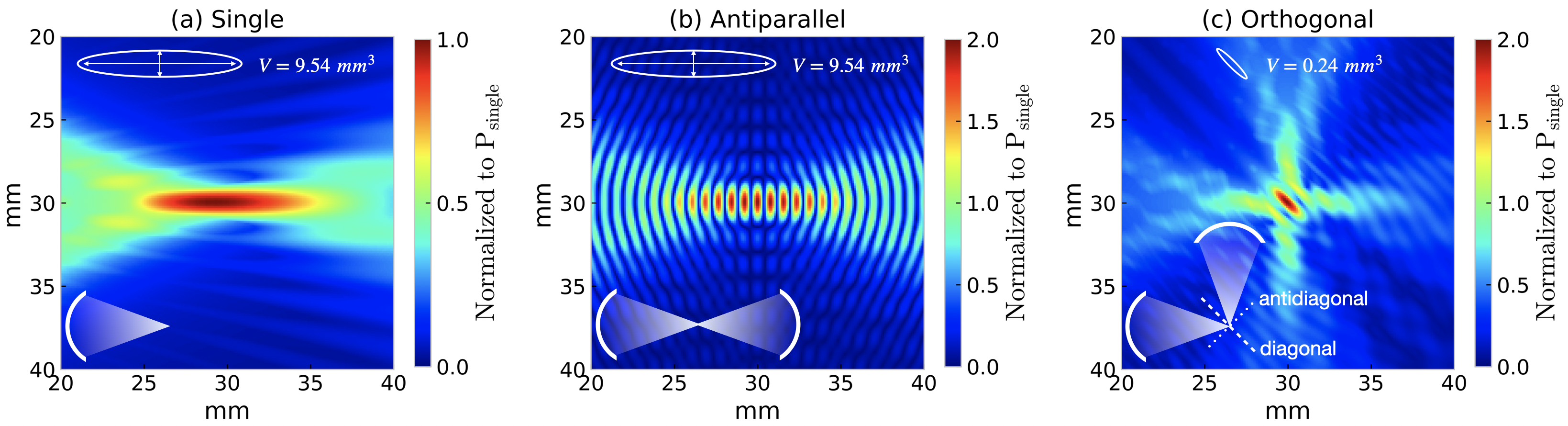}
\caption{Pressure amplitude profiles at 1 MHz in 2D for three different transducer configurations. Relative transducer positions are indicated in the lower left hand corner of each panel and all pressures are normalized to the single transducer. FWHM ellipsoid target volumes are shown in the top portion of each figure. \textbf{a}) A single transducer with its characteristic on-axis ellipsoid target. FWHM target volume was $9.54\ mm^3$. \textbf{b}) Antiparallel pair of transducers with on-axis standing waves. FWHM target volume was $9.54\ mm^3$. \textbf{c}) Orthogonal pair of transducers with the ellipsoid target rotated and aligned with the diagonal axis (dashed line) and the standing waves along the antidiagonal axis (dotted line). FWHM target volume was $0.24\ mm^3$.}
\label{fig:naftc3}
\end{figure*}

The exact gain in spatial resolution depends on the specifications of the transducers. However, optimal spatial resolution is achieved with an orthogonal pair of transducers compared to an antiparallel pair or a single transducer of identical specifications. In Fig. \ref{fig:naftc3}, we show the pressure amplitude profiles in 2D space across the single, antiparallel, and orthogonal transducer setups. Pressures were normalized to that of a single transducer to show the $2\times$ gain in pressure when two transducers were employed with overlapping foci, regardless of their spatial arrangements. 
Significant improvement in spatial resolution is evident from the size of the target in Fig. \ref{fig:naftc3}(c) compared to Fig. \ref{fig:naftc3}(a) and Fig. \ref{fig:naftc3}(b).

\begin{figure*}[ht]
\centering
\includegraphics[width=\textwidth]{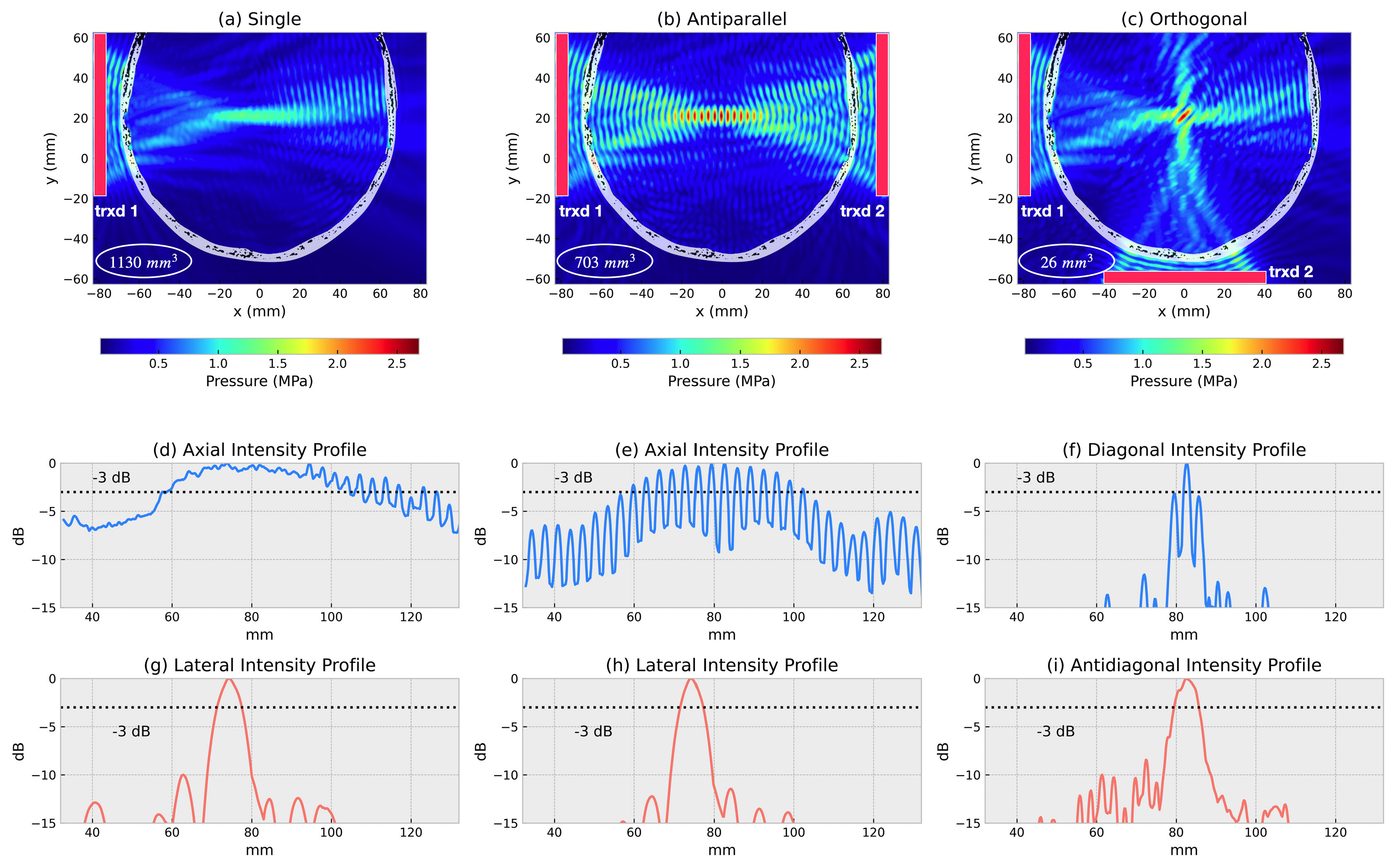}
\caption{Transcranial pressure amplitude and intensity profiles (rows) at 250 kHz for three different transducer configurations (columns). 84-mm wide, 420-element phased array transducers are shown in fuchsia. FWHM target volumes are displayed as insets at the lower left hand of the 2D pressure profiles. \textbf{a}) A single transducer pressure profile, accompanied by reflections off the skull, with a FWHM target volume of $1130\ mm^3$. \textbf{b}) Antiparallel setup generated standing waves along the axis of the two transducers with a FWHM target volume of $703\ mm^3$. \textbf{c}) Highly localized pressure profile of
the orthogonal arrangement with a FWHM target volume of $26\ mm^3$. \textbf{d-e}) Axial intensity profiles of the single and antiparallel setups, respectively. \textbf{f}) Diagonal intensity profile of the orthogonal configuration with its two side lobes below the -3 dB cutoff. \textbf{g-h}) Lateral intensity profiles of the single and antiparallel configurations, respectively. \textbf{i}) Antidiagonal intensity profile of the orthogonal setup.}
\label{fig:naftc4}
\end{figure*}

To quantitatively analyze the improvement in spatial resolution, we need to compute the target volumes in these three transducer configurations. Assuming target foci are in the shape of ellipsoids, obtaining their full width at half maximum (FWHM) of intensity volumes requires computing the major and minor axes of the ellipsoid at the $-3$ dB level of the intensity. Under these considerations, the target volume was $9.54\ mm^3$ for the single transducer, $9.54\ mm^3$ for the antiparallel transducers, and $0.24\ mm^3$ for the orthogonal transducers. For a pair of transducers, a simple $90$-degree rotation of one transducer resulted in a $39.75$ fold improvement in spatial resolution. These results, shown with Fig. \ref{fig:naftc3}, confirm that the antiparallel setup did not confer any advantages in terms of spatial resolution, because the broad axial beam profiles of the two transducers overlapped, resulting in yet another broad beam profile. The orthogonal setup leverages the relative sharpness of the lateral beam profile compared to the axial beam profile, and trims the axial beam profile of one transducer with the lateral beam profile of the other transducer and vice versa. 

This significant gain in spatial resolution holds in transcranial settings as well. In Fig. \ref{fig:naftc4}(a)-(c), we show the three configurations in the presence of a typical skull segment from a CT. The FWHM target volumes are shown as insets at the lower left hand of the figures. At $250$ kHz, we observed a FWHM target volume of $1130\ mm^3$ for the single transducer setup (Fig. \ref{fig:naftc4}(a)), and a FWHM target volume of only $26\ mm^3$ for the orthogonal configuration (Fig. \ref{fig:naftc4}(c)), corresponding to a 43-fold gain in spatial resolution. Fig. \ref{fig:naftc4}(d)-(e) show the axial intensity profiles for the single and antiparallel configurations, and Fig. \ref{fig:naftc4}(f) shows the diagonal intensity profile for the orthogonal setup. Whether the two side lobes in the diagonal intensity profile surpass the $-3$ dB threshold or not, depends on the transducer specifications and the simulation setup; the higher the focusing power of the transducers, the lower the side lobes. In Fig. \ref{fig:naftc4}(g)-(h) we show the lateral intensity profiles for the single and antiparallel setups, and in Fig. \ref{fig:naftc4}(i) we show the intensity profile for the antidiagonal axis of the orthogonal configuration.

\subsection{Acoustic Radiation Force in a Single Transducer}
\subsubsection{Complete Form}
\begin{figure*}[ht]
\centering
\includegraphics[width=\textwidth]{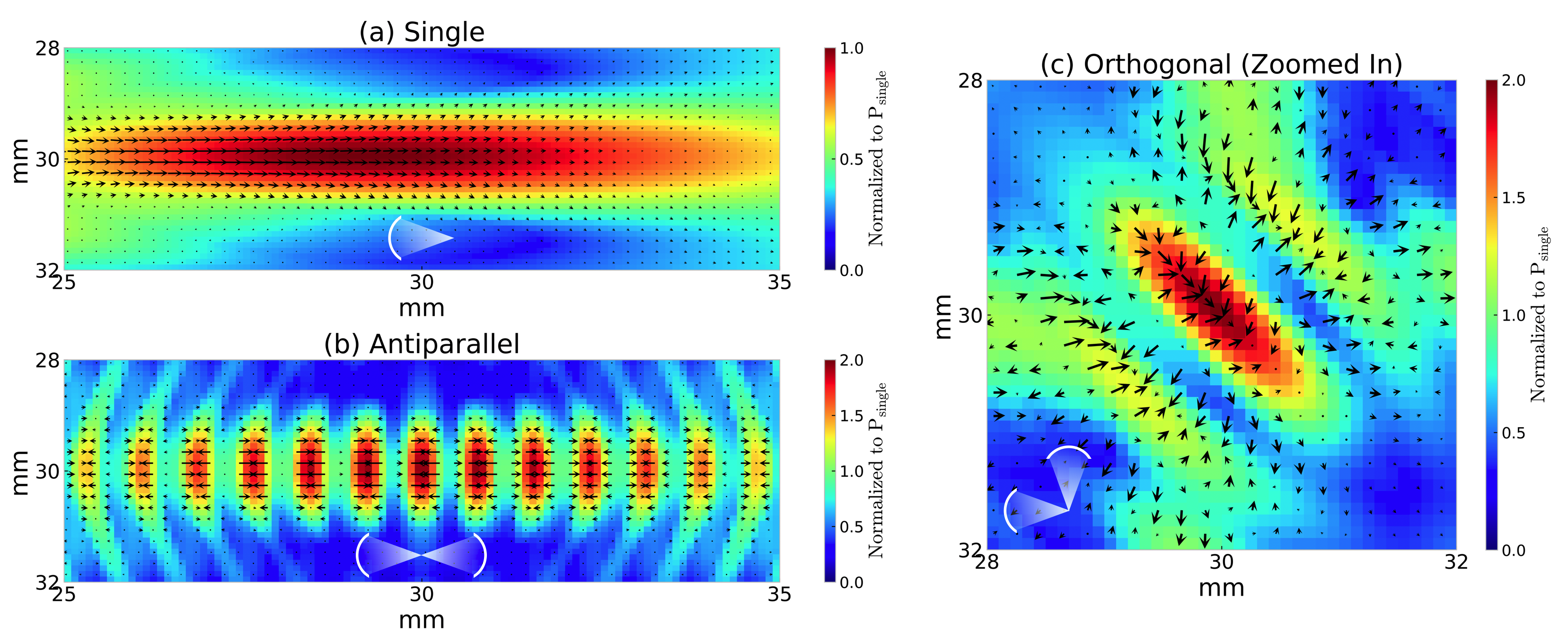}
\caption{ARF, computed with the complete solution, superimposed over the pressure field across the three transducer configurations. Transducer orientations are shown with the white insets in each figure. \textbf{a}) ARF vectors in a single transducer were unidirectional and pointed away from the transducer surface. They also reached their maximum value near the focus. \textbf{b}) Periodic ARF sinks and sources lied along the axis of the antiparallel configuration. At the nodes we observed compression, while at the antinodes we observed tension. \textbf{c}) Highly localized ARF sinks and sources were observed in the orthogonal regime. These regions of high compression and high tension, respectively, were confined to a $4\times 4\ mm^2$ area.}
\label{fig:naftc6}
\end{figure*}

The complete form of the acoustic radiation force for a single transducer was computed using equation (\ref{eq:ARF_complete}). To demonstrate the overall direction of the force in comparison to the pressure field, we superimposed the force vectors over the pressure amplitude for a single transducer (Fig. \ref{fig:naftc5}(a) and \ref{fig:naftc6}(a)). Our simulation of ARF in complete form was in agreement with previous valid simulations \cite{Prieur2016Simulationk-Wave} and Magnetic Resonance Acoustic Radiation Force Imaging (MR-ARFI) experiments \cite{McDannold2008MagneticImaging}.

\subsubsection{Quasi-Planar Wave Assumption}
\begin{figure*}[ht]
\centering
\includegraphics[width=\textwidth]{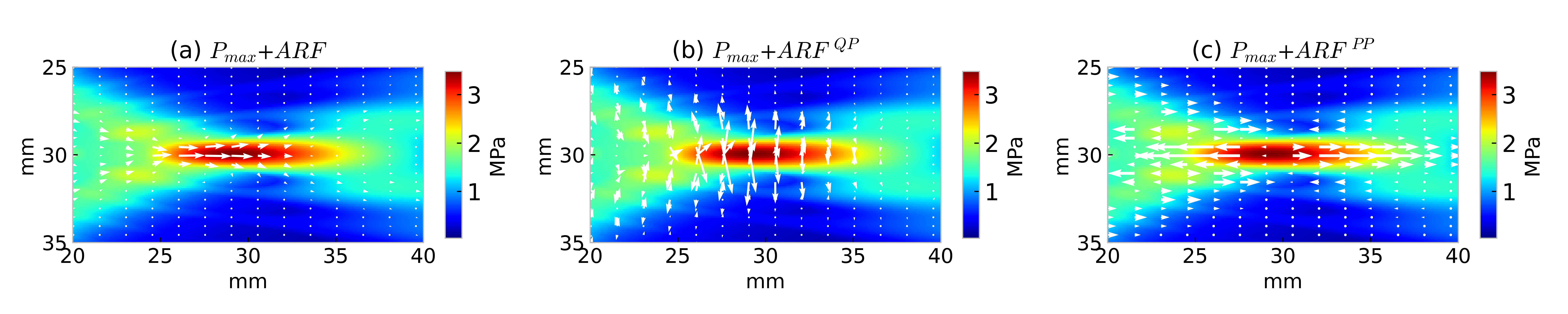}
\caption{ARF, computed with the complete solution and two tiers of simplifying assumptions, superimposed over the pressure field. \textbf{a}) ARF vectors reach their peak values about the focus and point away from the transducer surface. This is in agreement with MR-ARFI experiments. \textbf{b}) Under the qusi-planar wave assumption, the lateral components of ARF dominate the axial components, which is at odds with experimental observations. \textbf{c}) A purely-planar wave assumption neglects any lateral components and suggests that prior to the focus at the $30\ mm$ mark, there are forces pointing towards the transducer surface, which is entirely incorrect.}
\label{fig:naftc5}
\end{figure*}

Imposition of the quasi-planar wave assumption ignores the mean Eulerian excess pressure as well as its gradients. As shown in Fig. \ref{fig:naftc5}(b), under this tier of simplifying assumptions, when compared to the complete form, there was a gross exaggeration of the lateral ARF component relative to the axial ARF component, which defied both intuition and past experimental results \cite{McDannold2008MagneticImaging}.

\subsubsection{Purely-Planar Wave Assumption}
The simplest and at the same time the most limiting tier of these simplifying assumptions is the assumption of purely planar waves propagating along the transducer axis (Fig. \ref{fig:naftc5}(c)). Whereas the quasi-planar wave assumption overly exaggerated the lateral component of ARF, the purely-planar wave assumption took it to the opposite extreme and completely ignored it. More so than its value being incorrect, purely-planar ARF suggested there were on-axis forces towards the surface of the transducer immediately prior to the focus, which was categorically false. Forces should be pointing away from the transducer surface and any indication that some major attractive forces (pointing towards the transducer) were present in the ARF field would be at odds with previous studies \cite{McDannold2008MagneticImaging, Prieur2016Simulationk-Wave, Prieur2017ModelingElastography}.

\subsection{Acoustic Radiation Force in Antiparallel Configuration}
Close examination of the simplifying assumptions in computing ARF and the erroneous results they gave rise to, convinced us to use the complete form, equation (\ref{eq:ARF_complete}), in computing the ARF for the more complicated antiparallel and orthogonal configurations. In Fig. \ref{fig:naftc6}(b) we show the acoustic radiation force vectors over the pressure field of the antiparallel transducers. Note that in contrast to the unidirectionality of the force vectors in a single transducer (Fig. \ref{fig:naftc6}(a)), pointing away from the transducer surface (left to right), the antiparallel setup generated periodic sinks and sources, arranged along the axis of the two transducers. At the antinodes of the standing wave the forces were convergent (sinks) and at the nodes of the standing wave the forces were divergent (source). This interesting pattern, however, was not localized and spanned the entirety of the $10\ mm$-FWHM of the antiparallel configuration.

\subsection{Acoustic Radiation Force in Orthogonal Configuration}
\begin{figure*}[ht]
\centering
\includegraphics[width=\textwidth]{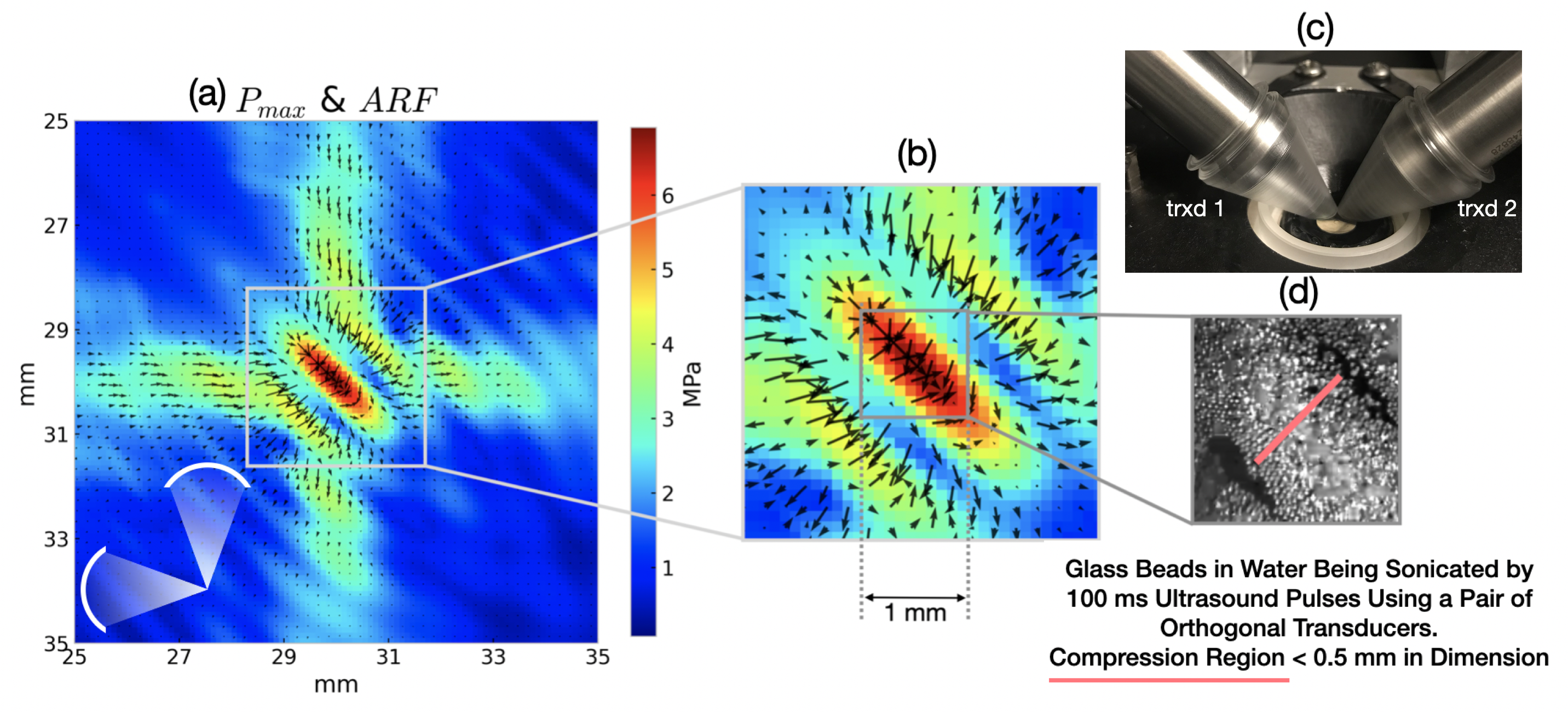}
\caption{ARF in orthogonal regime and its experimental validation. \textbf{a}) Pressure amplitude in orthogonal setup in MPa. The orientation of the two transducers is shown at the bottom left of the figure. \textbf{b}) A closer view of the force field shows the convergence of ARF at the focus and the divergence of ARF surrounding the focus. \textbf{c}) The orientation of the transducers in the experiment. Each transducer was elevated from the horizontal surface by $45^{\circ}$, forming a $90^{\circ}$ angle with one another. \textbf{d}) 50-micron glass beads suspended in water were pushed together at the focus, and they were pulled apart in regions surrounding the focus, as evidenced by the arrangement of the glass beads in space. This experiment corroborated the ARF simulations in \textbf{a} and \textbf{b}.}
\label{fig:naftc7} 
\end{figure*}

In Fig. \ref{fig:naftc6}(c) and Fig. \ref{fig:naftc7}(a)-(b) we show the pressure amplitude of the orthogonal setup, as well as the ARF vectors. Similar to the antiparallel configuration (Fig. \ref{fig:naftc6}(b)), ARF sinks and sources were periodically aligned. However, in contrast to the antiparallel configuration, these regions were along the antidiagonal axis of the transducers, and were confined to an area of only $4 \times 4\ mm^2$, instead of $4\times 10\ mm^2$. This suggested that if we used the orthogonal arrangement to sonicate a neural sub-population, those at the target (an ARF sink) would experience a sustained compression, while those immediately surrounding the target (ARF sources) would experience a sustained tension. To further test the validity of these simulations and be able to visually inspect the directionality of ARF in the orthogonal setup, we sonicated glass beads of approximately $50$ microns in diameter, suspended in water (Fig. \ref{fig:naftc7}(c)). As predicted by our simulations, the glass beads were pushed towards the focus (ARF sink), and pulled apart immediately surrounding the focus (where the ARF sources reside). As a result, there was a concentration of glass beads at the focus and an absence of glass beads surrounding the focus (Fig. \ref{fig:naftc7}(d)).

\newpage
\subsection{Unipolar Pressures}
\begin{figure*}[ht]
\centering
\includegraphics[width=\linewidth]{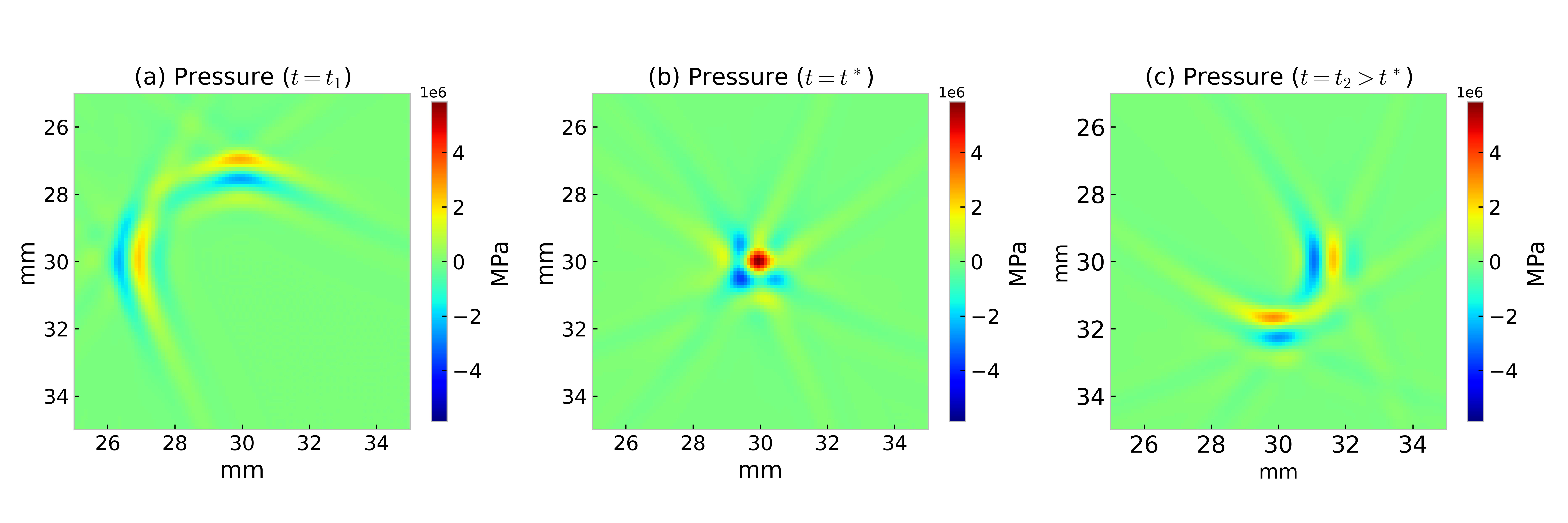}
\caption{The unipolar pressure method in the orthogonal setup. \textbf{a}) The tri-cycle pulses approaching each other at the ($30\ mm$, $30\ mm$) focus with opposite polarities. \textbf{b}) At time $t^*$, constructive superposition of the positive polarities of the pulses at the focus result in a \nicefrac{PPP}{PNP} of 1.54 over a FW65M surface area of $0.16\ mm^2$. \textbf{c}) Pulses move past each other in singular form, without any constructive or destructive interference that meets the FW65M cutoff.}
\label{fig:naftc8}
\end{figure*}

Tricycle pulses from two orthogonally arranged transducers were generated to demonstrate the unipolar pressure method. Once the pulses from the two transducers were inverted with respect to one another, we computed the appropriate time delay using the straightforward iterative approach presented in Algorithm \ref{alg:unipolar}. An optimal time delay was one that maximized the peak positive pressure to peak negative pressure ratio. Inversion and a 20-simulation timestep delay (as calculated via Algorithm \ref{alg:unipolar}) resulted in a dominant positive pressure, with a peak positive to peak negative pressure ratio of 1.54. This time delay needs to be determined accurately, otherwise the superposition may not result in a unipolar pressure. The unipolar positive pressure peaked at 5.9 MPa, with its base at 0 MPa spanning over 0.44 microseconds, about half the period. The unipolar pressure rise and fall demonstrated no nonlinearities and sudden rise in pressure resembling those used in histotripsy \cite{Duryea2011InHistotripsy, Roberts2005FocusedDirections, Roberts2014DevelopmentDirections, Xu2004ControlledErosion}, suggesting this approach could potentially be used as a safe method for unipolar neuromodulation with selective pressure polarity. In Fig. \ref{fig:naftc8} we show the 2D pressure field in the unipolar pressure method as a function of time. At time $t^*$ the two positive polarities superimpose and generate a dominant positive pressure. Note that $t^*$ denotes the time at which unipolar superposition takes place. There is no significance to its value as it will vary with the setup and the optimal time delay. At full width at 65\% of max pressure (FW65M, $-3.74$ dB), only the dominant positive pressure was observed while the negative pressure polarities fell under the FW65M cutoff. At this threshold, $x$ and $y$ dimensions of the peak positive pressure (Fig. \ref{fig:naftc8}(b)) measured at $0.4\ mm$ each, resulting in a surface area of $0.16\ mm^2$ that would experience such selective \nicefrac{PPP}{PNP} of 1.54 using two 1 MHz transducers in the orthogonal arrangement.

\subsection{Strain in a Single Transducer}
\subsubsection{Strain due to Particle Motion}
\begin{figure*}[ht!]
\centering
\includegraphics[width=\linewidth]{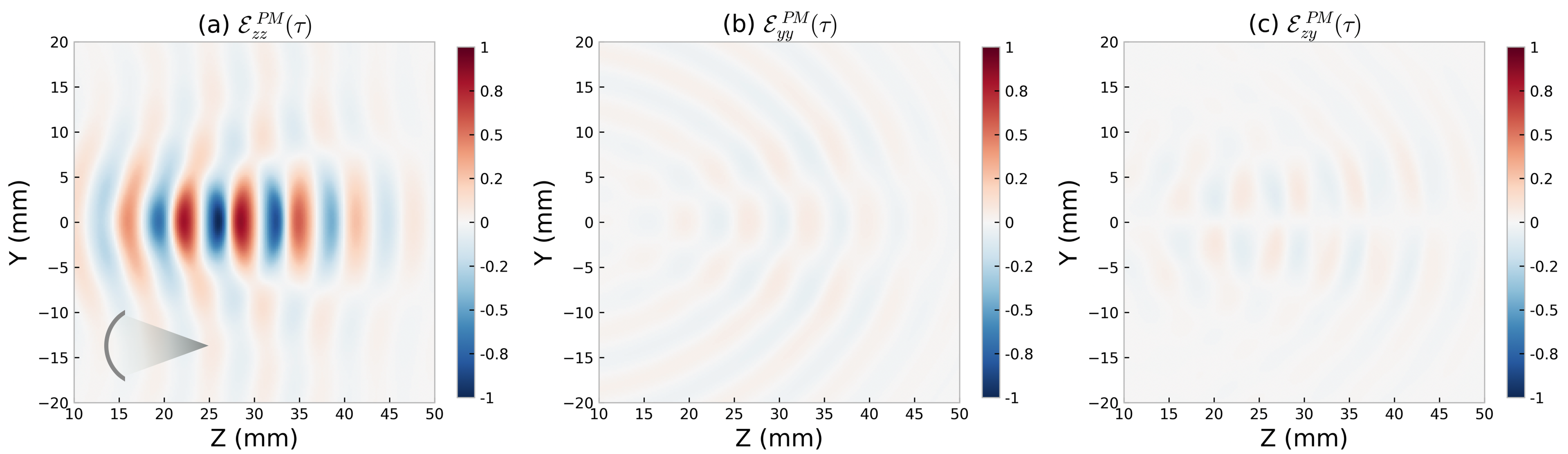}
\caption{Single transducer strain due to particle motion. Figures are normalized to the global maximum and minimum strain values across the three different strain components. \textbf{a}) Normal axial strain. Transducer location is shown by the inset at the lower left of the figure. Because on-axis particle oscillations had the largest amplitude at the focus, we observed that normal axial strain dominated the other strain components. \textbf{b}) Normal lateral strain. Particles oscillated both axially and laterally. However, axial oscillations dominated lateral oscillations and therefore, normal lateral strain had a lower value compared to normal axial strain. \textbf{c}) Shear strain. Shear strain due to particle motion was almost zero in the entire simulation field.}
\label{fig:naftc9}
\end{figure*}

Ultrasound consists of compression waves. Therefore, the largest particle oscillation amplitude will be along the axial dimension (the direction of wave propagation). Accordingly, as shown in Fig. \ref{fig:naftc9}, the dominant component of particle motion strain was the normal axial strain, $\mathcal{E}_{zz}^{PM}$ (Fig. \ref{fig:naftc9}(a)). The normal lateral component of $\mathcal{E}^{PM}$ (Fig. \ref{fig:naftc9}(b)) was significantly smaller, and the shear strain (Fig. \ref{fig:naftc9}(c)) was nearly zero. As the pulse traveled along the $z$-axis, the sign of strain varied between $+$ and $-$. In Fig. \ref{fig:naftc15}(a) we account for this variation in time and show the amplitude envelope over the dominant, normal axial component of particle motion strain. $\mathcal{E}^{PM}_{zz}$ reached its maximum value at the focus, and similar to the pressure profile, tapered off with distance from the focus.

\subsubsection{Strain due to ARF}
Strain under the effect of bulk tissue displacement due to ARF showed interesting properties in all its three components. We see in Fig. \ref{fig:naftc10}(a) that normal axial strain was positive prior to the focus ($z<30\ mm$). In that region tissue was stretched, under the effect of ARF. Conversely, past the focus ($z>30\ mm$), strain was negative, suggesting compression in that region of the tissue. At the focus however, normal axial strain remained zero. Normal lateral strain, Fig. \ref{fig:naftc10}(b), suggested that at the focus, along the transducer axis ($z=30\ mm$, $y=0\ mm$), tissue was compressed, while on either side of the transducer axis it was stretched out. The most interesting pattern was observed in the shear component of ARF strain, $\mathcal{E}^{ARF}_{zy}$. As shown in Fig. \ref{fig:naftc10}(c), the dominant component of ARF strain was the shear strain and it remained null at the focus, while it reached its peak values surrounding the focus. This stems from the fact that strain is the gradient of displacement, and that maximum tissue displacement due to ARF occurs at the focus.

\subsection{Strain in Antiparallel Configuration}
\subsubsection{Strain due to Particle Motion}
\begin{figure*}[ht]
\centering
\includegraphics[width=\linewidth]{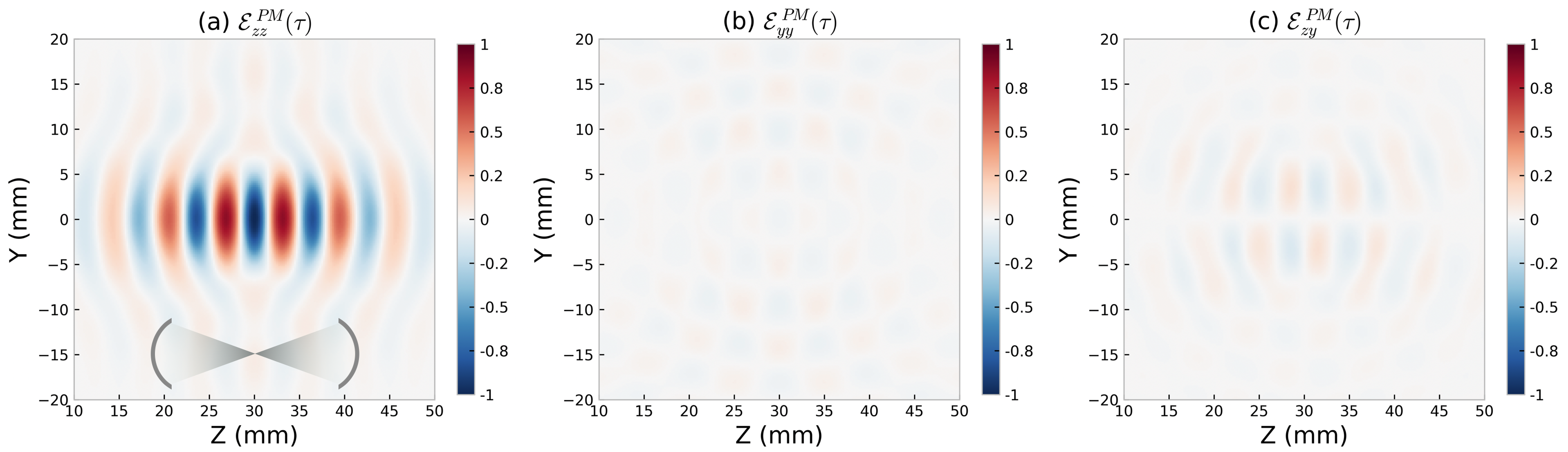}
\caption{Antiparallel configuration strain due to particle motion. Figures are normalized to the global maximum and minimum strain values across the three different strain components. \textbf{a}) Normal axial strain. The orientation of the transducers is shown by the inset at the bottom of the figure. We observed that normal axial strain dominated the other strain components. Moreover, given that two counter-propagating waves generate standing waves, strain at the nodes of the standing wave was zero and reached its maximum values at the antinodes. \textbf{b}) Normal lateral strain. Due to the curvature of the transducers, there were lateral components to tissue displacement. However, this normal lateral strain was significantly smaller than the normal axial strain. \textbf{c}) Shear strain. Shear strain due to particle motion was nearly zero.}
\label{fig:naftc11}
\end{figure*}

In the antiparallel configuration, $\mathcal{E}_{zz}^{PM}$ (Fig. \ref{fig:naftc11}(a)) dominated the other components of particle motion strain (Fig. \ref{fig:naftc11}(b)-(c)). This effect, similar to the case of a single transducer, was due to the fact that large-amplitude particle oscillations were along the axis of the two transducers. Because the antiparallel configuration gives rise to standing waves, strain at the nodes of the standing waves was zero. At the antinodes, strain was maximum and its sign alternated with time. 

\subsubsection{Strain due to ARF}
\begin{figure*}[ht]
\centering
\includegraphics[width=\linewidth]{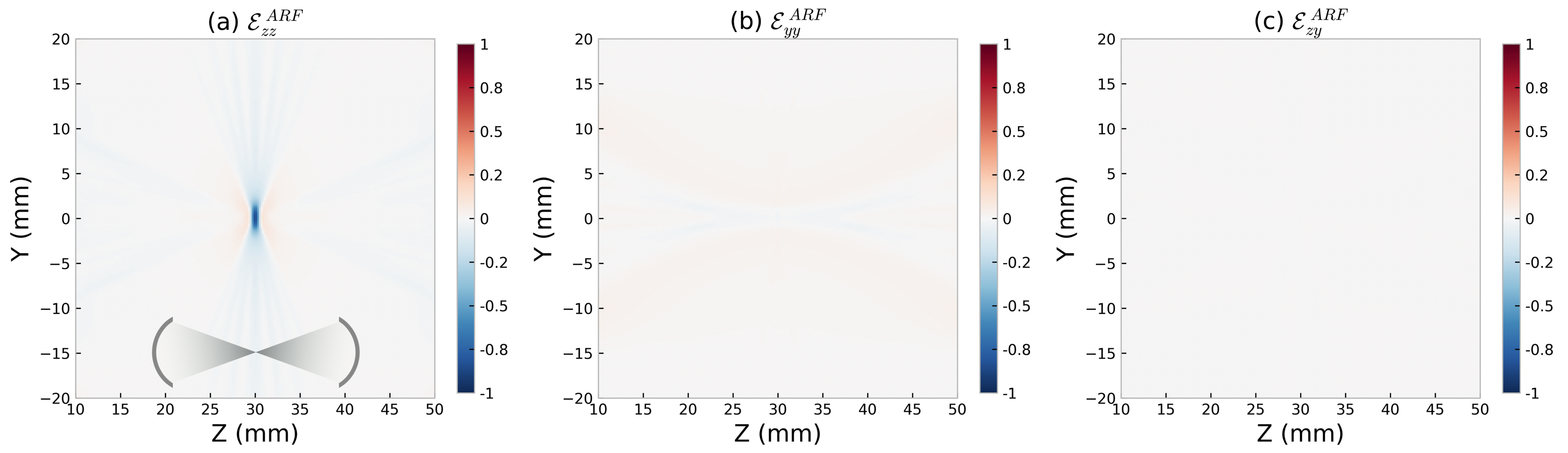}
\caption{Antiparallel configuration strain due to ARF. Figures are normalized to the global maximum and minimum strain values across the three different strain components. \textbf{a}) Normal axial strain. The orientation of the transducers is shown by the inset at the bottom of the figure. Because ARF from the two transducers pushed the tissue from both sides, we observed ``sandwiching'' of the tissue at the geometric center of the simulation setup (sharp negative strain). \textbf{b}) Normal lateral strain. Normal lateral strain was at least an order of magnitude smaller than the normal axial strain. \textbf{c}) Shear strain. The counter-propagating tissue displacement due to ARF resulted in minimal shear strain compared to the normal axial strain. In all of the figures, minor ripples in the images are due to simulation artifacts.}
\label{fig:naftc12}
\end{figure*}

Normal axial strain in this context dominated the other components of strain. As shown in Fig. \ref{fig:naftc12}(a), under the effect of ARF emanating from each transducer surface, the tissue was ``sandwiched'' at the focus, as evidenced by the sharp negative strain at the geometric focus of the setup. Both the normal lateral strain as well as shear strain (Fig. \ref{fig:naftc12}(b)-(c)) were insignificant in comparison to the normal axial strain. The high frequency ripples and ringing in Fig. \ref{fig:naftc12} are simulation artifacts and shall be ignored.

\subsection{Strain in Orthogonal Configuration}
\subsubsection{Strain due to Particle Motion}
\begin{figure*}[ht]
\centering
\includegraphics[width=\linewidth]{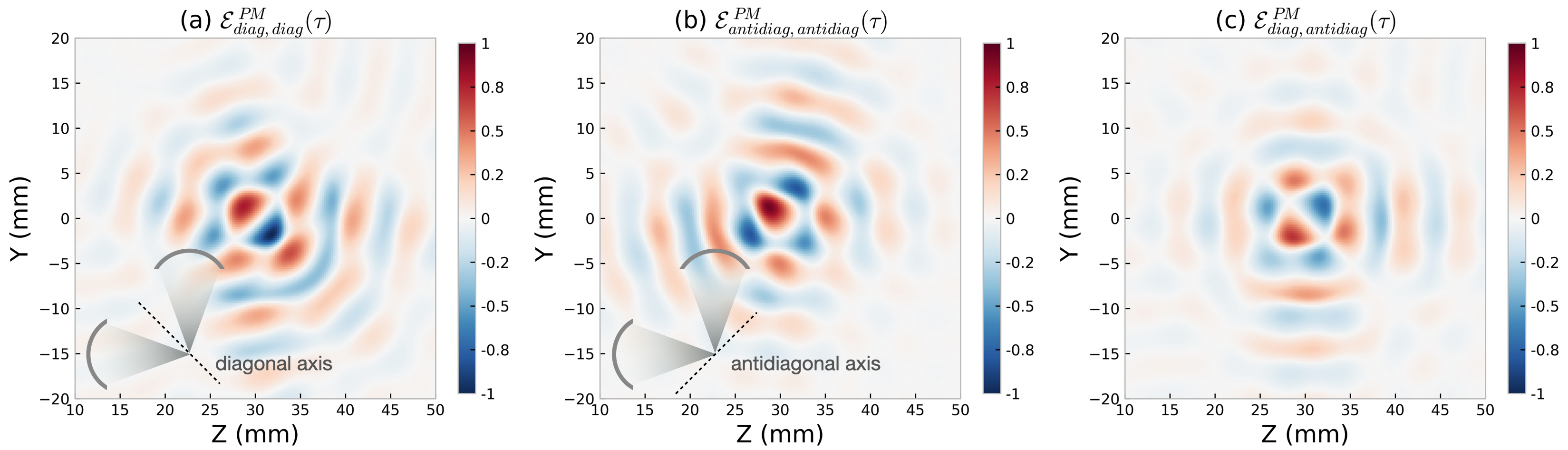}
\caption{Orthogonal configuration strain due to particle motion. Figures are normalized to the global maximum and minimum strain values across the three different strain components. Transducer locations, as well as the diagonal and antidiagonal axes are shown by the inset at the lower left of the figure. \textbf{a}) Normal diagonal strain. \textbf{b}) Normal antidiagonal strain. \textbf{c}) Shear strain. Because particle oscillations were equally decomposed and projected over the diagonal and antidiagonal axes, normal strains along these axes dominated the shear strain. Moreover, given the relative sharpness of the antidiagonal displacement profile compared to the diagonal displacement profile, the overall normal antidiagonal particle-motion strain (\textbf{b}) was greater than the normal diagonal strain (\textbf{a}).}
\label{fig:naftc13}
\end{figure*}

In the orthogonal configuration, propagating waves lie on the diagonal axis, and standing waves lie on the antidiagonal axis of the two transducers (Fig. \ref{fig:naftc3}(c)). Therefore, while we used equation (\ref{eq:strain}), instead of computing strain along the axial and lateral dimensions, we computed strain along the diagonal and antidiagonal directions. As evidenced by Fig. \ref{fig:naftc13}, normal strain due to particle motion in the diagonal dimension (Fig. \ref{fig:naftc13}(a)) was comparable to normal strain in the antidiagonal dimension (Fig. \ref{fig:naftc13}(ab), both of which dominated the shear strain (Fig. \ref{fig:naftc13}(c)). Antidiagonal strain was slightly greater than the diagonal strain due to its sharper displacement profile relative to the diagonal displacement profile. 

\subsubsection{Strain due to ARF}
\begin{figure*}[ht!]
\centering
\includegraphics[width=\linewidth]{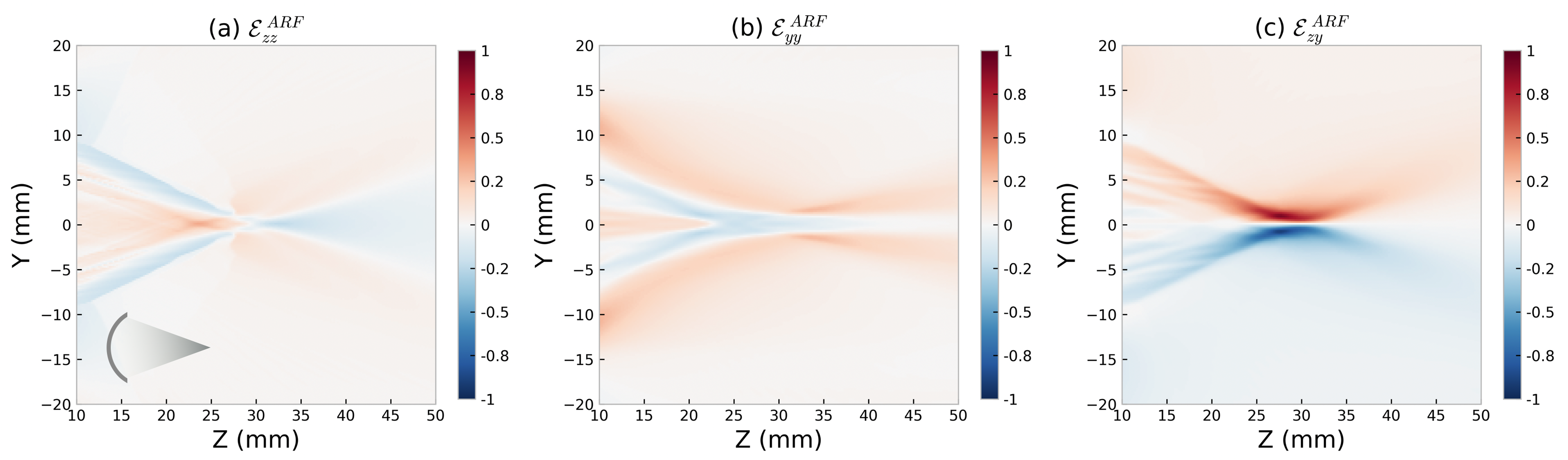}
\caption{Single transducer strain due to ARF. Figures are normalized to the global maximum and minimum strain values across the three different strain components. \textbf{a}) Normal axial strain. Transducer orientation is shown by the inset at the lower left of the figure. Because the tissue displacement reached its maximum around the focus, normal axial strain due to ARF was zero at the focus. \textbf{b}) Normal lateral strain. Along the transducer axis, normal lateral strain was negative, denoting compressive effects due to the curvature of the transducer. This was accompanied by positive normal strains on either side of the transducer axis. \textbf{c}) Shear strain. The dominant component of strain due to ARF was the shear strain. It reached its maximum around the focus and remained null at the focus.}
\label{fig:naftc10}
\end{figure*}

\begin{figure*}[ht]
\centering
\includegraphics[width=\linewidth]{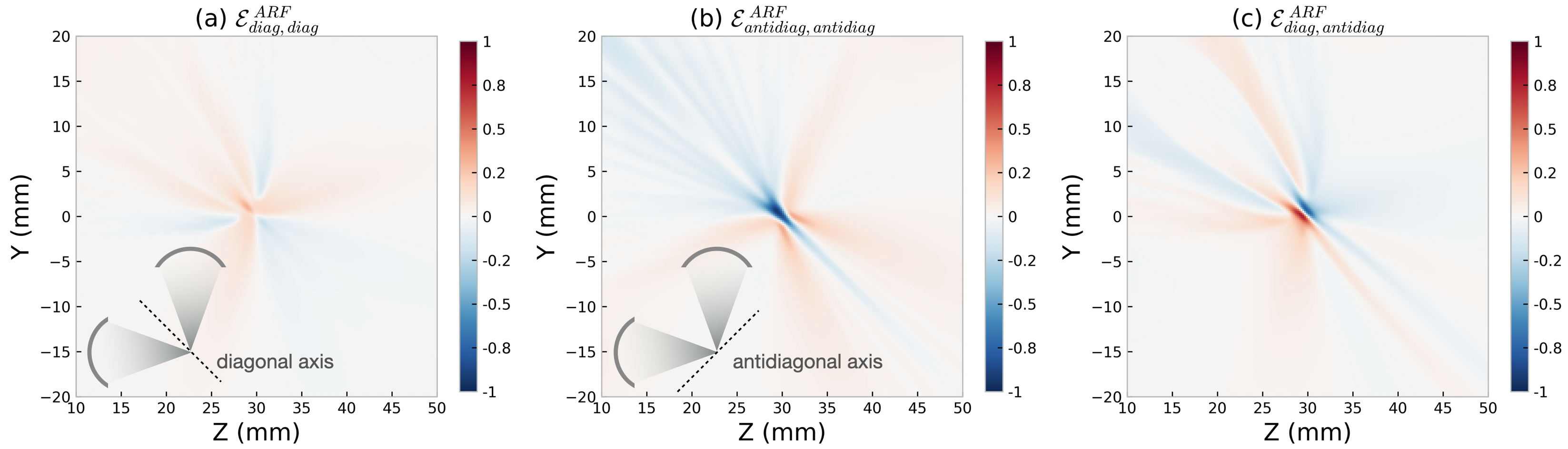}
\caption{Orthogonal configuration strain due to ARF. Figures are normalized to the global maximum and minimum strain values across the three different strain components. Transducer locations, as well as the diagonal and antidiagonal axes are shown by the inset at the lower left of the figure. \textbf{a}) Normal diagonal strain. We anticipated that normal diagonal strain would be zero at the focus, accompanied by tension (positive strain) and compression (negative strain) on either side of the focus along the diagonal axis. \textbf{b}) Normal antidiagonal strain. Due to the counter-propagating waves along the antidiagonal axis, tissue was compressed at the focus, as marked by the sharp negative strain at the $(30 mm, 30 mm)$ zone. \textbf{c}) Shear strain. Although shear strain was much greater than normal diagonal strain, it was still slightly less than the normal antidiagonal strain.}
\label{fig:naftc14}
\end{figure*}

Normal diagonal strain under ARF looked relatively similar to the normal axial strain of a single transducer (Fig. \ref{fig:naftc10}(a)), but rotated by $45^{\circ}$. At the geometric focus of the two transducers the normal diagonal strain was zero. As shown in Fig. \ref{fig:naftc14}, both the normal antidiagonal (Fig. \ref{fig:naftc14}(b)) as well as shear strains (Fig. \ref{fig:naftc14}(c)) dominated the normal diagonal strain (Fig. \ref{fig:naftc14}(a)). However, antidiagonal strain was the largest strain component, under the compressive effect of the counter-propagating waves from the two transducers. 

\begin{figure*}[ht!]
\centering
\includegraphics[width=\linewidth]{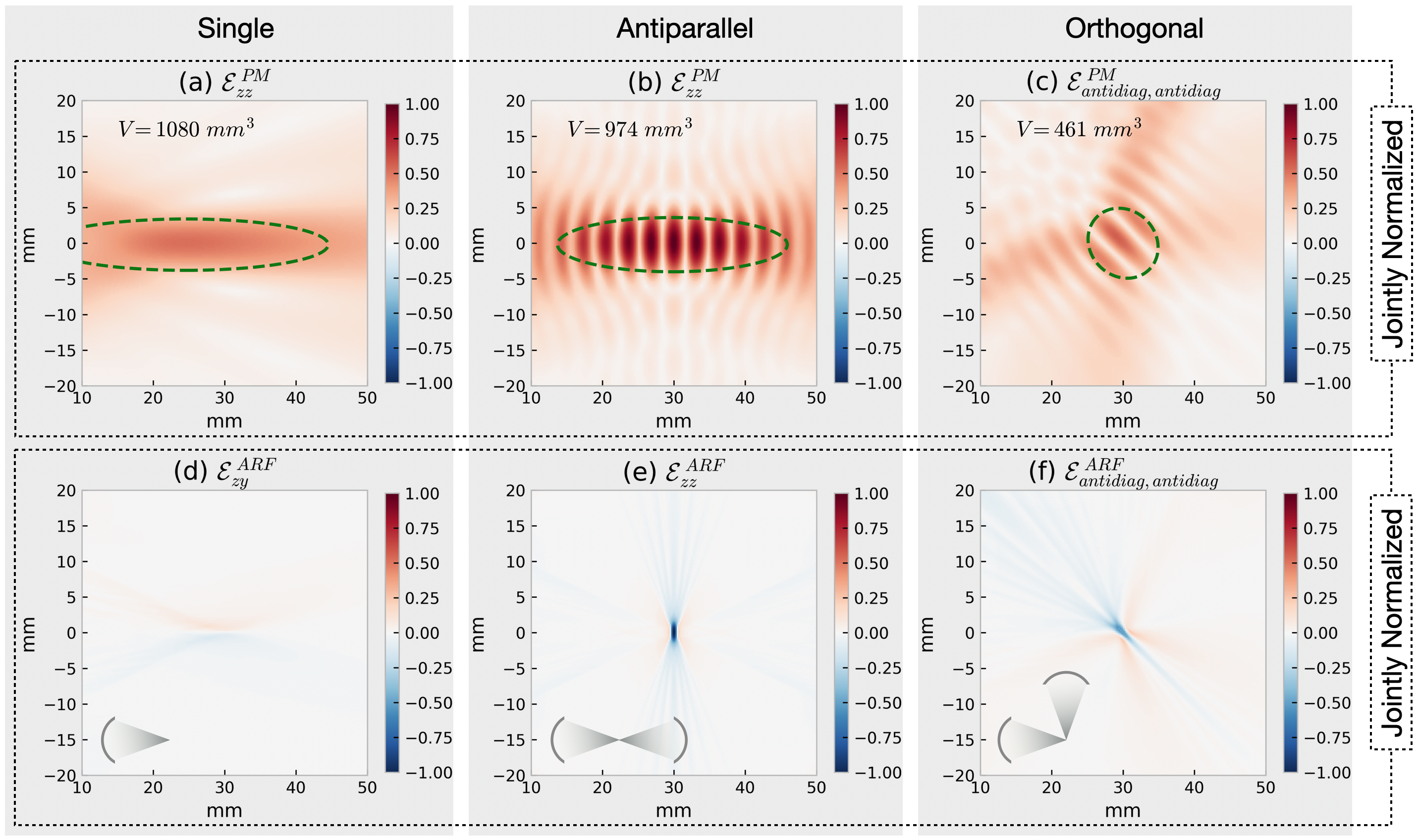}
\caption{Comparison of dominant components of strain due to particle motion (PM) and strain due to acoustic radiation force (ARF) across the three configurations. Top row represents PM strains, and bottom row represents ARF strains. Each row is normalized independently, to the global maximum and minimum strain values in that row. Each column represents one of the three configurations, denoted by the text at the top, and the graphical inset at the bottom of the columns. FWHM focal volumes for PM strain are shown with the ellipses in dashed green. Numerical values of the FWHM volumes are shown on the upper left of each figure. Note that these are 20-cycle pulses at 250 kHz. \textbf{a}) Normal axial strain due to PM in a single transducer. \textbf{b}) Normal axial strain due to PM in the antiparallel setup. \textbf{c}) Normal lateral strain due to PM in the orthogonal setup. \textbf{d}) Shear strain due to ARF in the single transducer configuration.  \textbf{e}) Normal axial strain due to ARF in the antiparallel configuration.  \textbf{f}) Normal lateral strain due to ARF in the orthogonal configuration. Magnitudes of strain, both due to PM as well as ARF, were a function of pressure. As such, the antiparallel setup had the largest strain (PM and ARF) values, followed by the orthogonal setup, and the single transducer setup. In terms of resolution of the PM strain, the orthogonal configuration in (\textbf{c}) was confined to a smaller area in comparison to the other two configurations.}
\label{fig:naftc15}
\end{figure*}

In Fig. \ref{fig:naftc15}, we display the peak component of particle motion strain and the peak component of ARF strain for each of the three configurations. All particle motion strains were jointly normalized (Fig. \ref{fig:naftc15}(a)-(c)) to their global maximum value, and all ARF strains were normalized together (Fig. \ref{fig:naftc15}(d)-(f)) to their global maximum values. As a result, the top row figures can be compared against one another, the bottom row figures can also be compared against one another, but the normalized amplitudes of the top row figures shall not be compared with the normalized amplitudes of the bottom row figures. The antiparallel configuration had the largest strain value for both particle motion (Fig. \ref{fig:naftc15}(b)) and ARF (Fig. \ref{fig:naftc15}(e)). This was due to the greater particle and tissue displacements in the antiparallel setup, which resulted from higher pressures generated when two transducers faced each other. Following the same logic, the strain values for the orthogonal configuration (Fig. \ref{fig:naftc15}(c) and Fig. \ref{fig:naftc15}(f)) and a single transducer (Fig. \ref{fig:naftc15}(a) and Fig. \ref{fig:naftc15}(d)) ranked second and third, respectively, in terms of decreasing strain magnitude. In terms of localization of strain, however, the normal particle motion strain in the orthogonal setup (Fig. \ref{fig:naftc15}(c)) was more focused compared to the other two configurations.

\section{Discussion}
The results of this study demonstrate that using two 1 MHz transducers in an orthogonal configuration can achieve targets as small as $0.24\ mm^3$ deep in the brain, an impressive 40-fold gain in spatial resolution compared to a single transducer with identical specifications. Importantly, such significant gains in spatial resolution hold for both high and low frequencies, and free water as well as transcranial settings, as we saw in the 1 MHz free water and 250 kHz transcranial simulations. However, the exact gain will remain a function of transducer specifications. Moreover, we showed that standing waves, identified as desirable in eliciting neural activity in salamander retina \cite{Menz2019RadiationRetina}, were generated along the antidiagonal axis of the orthogonal setup. By providing a thorough treatment of the acoustic radiation force (ARF), we revealed the shortcomings associated with different tiers of simplifying assumptions. Computation of ARF in the orthogonal setup, without the simplifying assumptions that break in the limit of focused ultrasound beams, showed highly localized regions of compression (at ARF sinks) and tension (at ARF sources) along the antidiagonal axis of the orthogonal setup. This regimented ARF field enables the study of neural sub-populations' responses to compression versus tension on a fine spatial scale. 

One potential disadvantage of using transducers positioned orthogonally compared to the antiparallel setup is efficiency. In the orthogonal setup, the vector components of ARF from each transducer have to be decomposed into their projections along the antidiagonal and diagonal axes. Because the transducers are at 90-degree angles, these projections are calculated by multiplying the vector components by $\cos\ 45^{\circ}$. The projected components onto the antidiagonal axis interact directly to generate standing waves along the antidiagonal axis, while the remaining components, projected onto the diagonal axis, interact as propagating waves. In the antiparallel setup, both transducers are facing each other and are positioned on-axis, allowing the entirety of their ARF vectors to interact directly without needing decomposition or projection, as is the case in the orthogonal setup. This direct interaction of the full vector components, rather than only their $\nicefrac{\sqrt{2}}{2}$ fractions, makes the antiparallel setup more efficient. However, the orthogonal arrangement outperforms the antiparallel arrangement substantially in terms of spatial resolution. As observed, a pair of antiparallel transducers with identical specifications, though capable of generating standing waves on-axis, did not yield any improvement in spatial resolution compared to a single transducer. Although Riis et al. \cite{Riis2022Multifrequency-basedVolume} demonstrated that frequency modulation sharpened the focal volume in the antiparallel setup, this approach comes with three drawbacks when compared to the orthogonal setup. First, the gain in spatial resolution is still less than with the orthogonal setup. Second, to generate the necessary destructive/constructive interference for suppression of off-target pressures, phased transducers are needed for the multi-frequency approach in the antiparallel setup, whereas the orthogonal arrangement can be used with either single-element or phased-array transducers. In the orthogonal setup, phased array transducers will only be needed in transcranial settings where aberrations due to the skull are significant and need to be corrected for each transducer to ensure constructive interference at the intended target. Third, in transcranial settings, with frequency modulation for the antiparallel setup, phase aberration corrections due to the skull have to simultaneously be accounted for, in addition to focal volume sharpening via frequency modulation. This turns the problem into a bi-objective one, with a trade-off between accurate energy deposition at the target and focal volume sharpening. In contrast, when orthogonal transducers are used transcranially, the sole objective is phase aberration correction to ensure constructive interference at the target; the 90-degree arrangement of the transducers automatically takes care of focal volume sharpening.

To further expand our toolbox in transcranial ultrasound stimulation (TUS), we introduced the unipolar method in the orthogonal setup, which allows for imparting dominant pressures of single polarity on the target so that the possible selective response of neurons to pressure polarity can be investigated. Application of unipolar pressures is not restricted to generating single dominant peaks. A variety of duty cycles and pulse repetition frequencies could be designed with appropriate delays to generate unipolar pulse trains. A unipolar method will require phase aberration correction if employed transcranially. In such scenarios, Algorithm \ref{alg:unipolar} will have to include phase correction for each iteration. Depending on the phase correction method, computational cost could be a potential drawback in transcranial unipolar method. 

The field has not yet taken a deep dive into the mechanical effects of TUS on neural tissue. Pressure, intensity, and pulsing scheme remain the dominant metrics reported across the board {\cite{King2013EffectiveNeurostimulation, Tufail2010TranscranialCircuits, Legon2014TranscranialHumans, Lee2015Image-GuidedCortex, Folloni2019ManipulationStimulation}}. However, we find it imperative to pair those metrics with known mechanical effects on the tissue, such as ARF and strain. Strain, as discussed, can arise from both particle motion and ARF. Although they are related and co-exist, they present vastly different strain maps. Interestingly, for a given pressure at the focus, strain magnitude due to particle motion did not change as a function of frequency. This was because the smaller displacement gradients at lower frequencies were offset by the larger particle displacements, and vice versa. We hypothesize that in the limit of higher frequencies, strain due to ARF will dominate strain due to particle motion for two reasons: First, ARF scales with tissue attenuation, and attenuation is greater at higher frequencies. Second, past a certain threshold, neural tissue will no longer be able to resolve the rapid particle oscillations and will perceive displacement due to particle motion as a constant signal, gradient of which is zero. Further analysis and experimental setups are needed to explore and validate these hypotheses. Unfortunately, we did not have the computational resources to explore the exact frequency ranges where one form of strain dominates the other. Therefore, we refrain from specifying a critical frequency.
Moreover, since we are in the continuum level, there are many neurons at any given point of the simulation frame. Because strain is a tensor, depending on the neuron direction with respect to the principal directions of this tensor, stretch or shear type deformation will be experienced by the neuron. Investigation of the orientation-selectivity of the neurons towards strain is outside the scope of this work and is currently being investigated.

In our quest to elucidate the mechanism of interaction in TUS and build a mapping between ultrasound parameters and neural activity, it is crucial to study neural subpopulations on a finer scale, probe their likely differential responses under divergent and convergent ARF, impart selective pressures on the target to explore potential selectivity to pressure, and understand the particle motion and ARF strain profiles of the region under investigation. In this manuscript, we have provided the theoretical tools necessary for conducting these experimental investigations.

\appendix
\subsection*{Divergence Theorem}
The divergence theorem allows for rewriting a volume integral over a continuously differentiable vector field \textbf{F}, as a surface integral over its boundary, and vice versa: 

\begin{equation}\label{eq:div_theorem}
    \iiint_V(\nabla \cdot \textbf{F})dV = \varoiint_S (\textbf{F}\cdot \hat{\textbf{n}})dS,
\end{equation}
where the LHS is the volume integral of the volume $V$, $\nabla \cdot$ is the divergence operator, and the RHS is the surface integral over the boundary of the volume $V$. $\textbf{F}\cdot\hat{\textbf{n}}$ is the component of the vector field perpendicular to the surface, denoted by $\hat{\textbf{n}}$. For more details on the divergence theorem, please refer to \emph{Div, Grad, Curl, and All That} by H. M Schey \cite{Schey2004DivCalculus}.

\subsection*{Gradient Theorem}
\begin{equation}\label{eq:one}
\begin{split}
    \vec{k}\cdot \nabla \Phi &= \big(k_1, k_2, \dots , k_n\big)^T \bigg( \frac{\partial\Phi}{\partial x_1}, \frac{\partial \Phi}{\partial x_2}, \dots, \frac{\partial\Phi}{\partial x_n}\bigg) \\
    &= \sum_{i=1}^n k_i\frac{\partial \Phi}{\partial x_i}.
\end{split}
\end{equation}

\begin{align}\label{eq:two}
\begin{split}
    \nabla \cdot (\Phi \vec{k}) &= \nabla \cdot(\Phi k_1, \Phi k_2, \dots, \Phi k_n) \\
    &= \frac{\partial}{\partial x_1}(\Phi k_1) + \frac{\partial}{\partial x_2}(\Phi k_2) + \dots + \frac{\partial}{\partial x_n}(\Phi k_n) \\
    &= k_1\frac{\partial \Phi}{\partial x_1} + \Phi \frac{\partial k_1}{\partial x_1} + k_2\frac{\partial \Phi}{\partial x_2} + \Phi \frac{\partial k_2}{\partial x_2} + \dots \\
    &+ k_n\frac{\partial \Phi}{\partial x_n} + \Phi \frac{\partial k_n}{\partial x_n}\\ 
    &= \sum_{i=1}^nk_i\frac{\partial \Phi}{x_i} + \Phi \sum_{i=1}^n\frac{\partial k_i}{\partial x_i}.
\end{split}
\end{align}

\begin{align}\label{eq:three}
\begin{split}
    \Phi \big( \nabla \cdot \vec{k}\big) &= \Phi \bigg( \frac{\partial k_1}{\partial x_1} + \frac{\partial k_2}{\partial x_2} + \dots + \frac{\partial k_n}{\partial x_n}   \bigg) \\ 
    &= \Phi \sum_{i=1}^n \frac{\partial k_i}{\partial x_i}.
\end{split}
\end{align}

Putting together equations \ref{eq:one}, \ref{eq:two}, and \ref{eq:three}, we have:

\begin{equation}\label{eq:useful_theorem}
    \vec{k}\cdot \nabla \Phi = \nabla \cdot (\Phi \vec{k}) - \Phi(\nabla \cdot \vec{k}).
\end{equation}

\begin{align}\label{eq:grad_theorem}
\begin{split}
    \vec{k}\cdot \iiint \nabla \Phi dV &= \iiint\vec{k}\cdot \nabla \Phi dV \\ 
    &= \iiint \bigg\{ \nabla \cdot (\Phi \vec{k}) - \Phi(\nabla \cdot \vec{k})\bigg\} dV  \text{\ \ from (\ref{eq:useful_theorem})} \\ 
    &= \iiint \nabla \cdot (\Phi \vec{k})dV - \iiint \Phi(\nabla \cdot \vec{k}) dV  \\ 
    &= \iiint \nabla \cdot (\Phi \vec{k})dV \text{\ \ since $\vec{k}$ is fixed}  \\ 
    &= \iint (\Phi \vec{k})\cdot \hat{\textbf{n}}dS \\
    &\text{\ \ invoking the divergence theorem} \\ 
    &= \vec{k}\cdot\iint \Phi \hat{\textbf{n}}dS \text{\ \ since $\vec{k}$ is fixed}\\ 
    \vec{k}\cdot \iiint \nabla \Phi dV &= \vec{k}\cdot\iint \Phi \hat{\textbf{n}}dS\\
    &\Rightarrow \\
    \iiint \nabla \Phi dV &= \iint \Phi \hat{\textbf{n}}dS,
\end{split}
\end{align}
 
\noindent which is the gradient theorem.

\subsection*{Acoustic Radiation Force}
Impulse of a force applied to a control volume results in change of momentum in the control volume:

\begin{align}\label{impulse_app}
\begin{split}
    d\textbf{P}_V &= \textbf{F}dt \\
    \frac{d\textbf{P}_V}{dt} &= \textbf{F},
\end{split}
\end{align}

\noindent whereby force vector \textbf{F} could be directly computed from the rate of change of momentum vector $\textbf{P}_V$ in the control volume of interest. The left-hand side (LHS) in equation (\ref{impulse_app}) could be written out explicitly as

\begin{equation}\label{momentum_flux_app}
    \frac{d\textbf{P}_V}{dt} = \frac{d\textbf{P}}{dt} + \frac{d\textbf{P}_{\text{out}}}{dt} - \frac{d\textbf{P}_{\text{in}}}{dt}.
\end{equation}

The right-hand side (RHS) requires some explanation \cite{MarkDrela2005FluidMechanics, JohnD.Anderson2001FundamentalsAerodynamics}: 

\begin{itemize}
    \item $\frac{d\textbf{P}}{dt}$: rate of change of instantaneous momentum inside the control volume, where $\textbf{P}(t)=\iiint\rho \textbf{v}dV$. This force component is equivalent to the ARF. 
    \item $\frac{d\textbf{P}_{\text{out}}}{dt}$: rate of momentum leaving the control volume due to mass flow.
    \item $\frac{d\textbf{P}_{\text{in}}}{dt}$: rate of momentum entering the control volume due to mass flow.
\end{itemize}

Substituting \textbf{F} for the LHS in equation (\ref{momentum_flux_app}), and recognizing that $\frac{d\textbf{P}_{\text{out}}}{dt} - \frac{d\textbf{P}_{\text{in}}}{dt}$ is momentum flux through a closed area $A$, that is $\varoiint \rho(\textbf{v}\cdot\hat{\textbf{n}})\textbf{v}dA$, we can rewrite equation (\ref{momentum_flux_app}) as:

\begin{equation}\label{momentum_flux_integrals_app}
    \textbf{F} = \frac{d}{dt}\iiint\rho \textbf{v}dV + \varoiint \rho(\textbf{v}\cdot\hat{\textbf{n}})\textbf{v}dA.
\end{equation}

We can expand the LHS into body forces, notably  the gravity, and surface forces, pressure and viscosity, to write out the integral momentum equation \cite{MarkDrela2005FluidMechanics, JohnD.Anderson2001FundamentalsAerodynamics}:

\begin{equation}\label{eq:integral_momentum_app}
\begin{split}
    \iiint \rho\textbf{g}dV - &\varoiint p\hat{\textbf{n}}dA + \textbf{F}_{\text{viscous}} = \\
    &\frac{d}{dt}\iiint\rho \textbf{v}dV + \varoiint \rho(\textbf{v}\cdot\hat{\textbf{n}})\textbf{v}dA,
\end{split}
\end{equation}

\noindent where the LHS terms are gravitational force, pressure force, and force due to viscosity. Since the timescale over which particle collisions take place is very short, we can ignore the gravitational force. Moreover, because the bulk modulus in ultrasound is much greater than the shear modulus, and that normal stresses are dominant compared to shear stresses, it is reasonable to drop the viscous force \cite{Prieur2017ModelingElastography}. Dropping these terms simplifies equation (\ref{eq:integral_momentum_app}): 

\begin{equation}\label{eq:integral_momentum_inviscid_app}
    \frac{d}{dt}\iiint\rho \textbf{v}dV = -\varoiint p\hat{\textbf{n}}dA - \varoiint\rho(\textbf{v}\cdot\hat{\textbf{n}})\textbf{v}dA,
\end{equation}

As we see in equation (\ref{eq:integral_momentum_inviscid_app}) momentum flux and force due to pressure are in the form of surface integrals, whereas ARF is a volume integral. In order to convert the momentum flux and pressure force to volume integrals, we will invoke the divergence theorem, equation (\ref{eq:div_theorem}), and gradient theorem, equation (\ref{eq:grad_theorem}), respectively: 

\begin{align}\label{eq:pressure_to_volume_app}
\begin{split}
    \varoiint p\hat{\textbf{n}}dA &= \iiint \nabla pdV \\
    &\text{\ \ following the gradient theorem in \ref{eq:grad_theorem}}.
\end{split}
\end{align}

\begin{align}\label{eq:momentum_to_volume_app}
\begin{split}
    \varoiint\rho(\textbf{v}\cdot\hat{\textbf{n}})\textbf{v}dA &= \varoiint \rho(\textbf{v}\cdot \hat{\textbf{n}}) (v_i\hat{i} + v_j\hat{j} + v_k\hat{k})dA \\
    &\text{\ \ since \textbf{v} is a vector}\\ 
    &= \varoiint \rho(\textbf{v}\cdot \hat{\textbf{n}}) v_idA\hat{i} + \varoiint \rho(\textbf{v}\cdot \hat{\textbf{n}}) v_jdA\hat{j} \\
    & + \varoiint \rho(\textbf{v}\cdot \hat{\textbf{n}}) v_kdA\hat{k} \\ 
    &= \iiint \nabla \cdot (\rho \textbf{v}v_i)dV\hat{i} + \iiint \nabla \cdot (\rho \textbf{v}v_j)dV\hat{j} \\
    &+ \iiint \nabla \cdot (\rho \textbf{v}v_k)dV\hat{k} \\
    &\text{\ \ using the divergence theorem in \ref{eq:div_theorem}}. \\ 
\end{split}
\end{align}

At this point we have converted all the surface integrals to volume integrals. The time derivative in the LHS in equation (\ref{eq:integral_momentum_inviscid_app}) could be taken inside the integral:

\begin{equation}\label{eq:lhs_arf_app}
    \frac{d}{dt}\iiint\rho \textbf{v}dV = \iiint\frac{d}{dt} (\rho \textbf{v})dV.
\end{equation}

Substituting equation (\ref{eq:pressure_to_volume_app}) and (\ref{eq:momentum_to_volume_app}) for the RHS in equation (\ref{eq:integral_momentum_inviscid_app}) yields:

\begin{equation}\label{eq:rhs_arf_app}
\begin{split}
    -\varoiint p\hat{\textbf{n}}dA &- \varoiint\rho(\textbf{v}\cdot\hat{\textbf{n}})\textbf{v}dA = \\
    &-\iiint \nabla pdV - \iiint \nabla \cdot (\rho \textbf{v}v_i)dV\hat{i} \\
    &- \iiint \nabla \cdot (\rho \textbf{v}v_j)dV\hat{j} - \iiint \nabla \cdot (\rho \textbf{v}v_k)dV\hat{k}.
\end{split}
\end{equation}

Putting them all together, we can rewrite equation (\ref{eq:integral_momentum_inviscid_app}) in terms of volume integrals only: 

\begin{align}
\begin{split}
    \iiint\frac{d}{dt} (\rho \textbf{v})dV &= -\iiint \nabla pdV - \iiint \nabla \cdot (\rho \textbf{v}v_i)dV\hat{i} \\
    &- \iiint \nabla \cdot (\rho \textbf{v}v_j)dV\hat{j} - \iiint \nabla \cdot (\rho \textbf{v}v_k)dV\hat{k}.
\end{split}
\end{align}

To simplify the math, let's focus only on the $i^{th}$ component: 
\begin{align}
\begin{split}
    \iiint\frac{d}{dt} (\rho v_i)dV &= -\iiint (\nabla p)_i dV - \iiint \nabla \cdot (\rho \textbf{v}v_i)dV\\ 
    &= -\iiint \bigg[(\nabla p)_i + \nabla\cdot (\rho \textbf{v}v_i) \bigg]dV.
\end{split}
\end{align}

Now that all of our integrals are volume integrals, we can approximate the volume integral over a control volume by replacing $\iiint (\cdot)dV$ with $\delta V$:

\begin{equation}
    \frac{d}{dt} (\rho v_i)\delta V = - \bigg[(\nabla p)_i + \nabla\cdot (\rho \textbf{v}v_i) \bigg]\delta V,
\end{equation}

\noindent where $\delta V$ terms can be dropped from either end. Recognizing the $\frac{d}{dt}(\rho v_i)$ as the $i^{th}$ component of ARF, we have:

\begin{align}\label{eq:arf_t_app}
    ARF(t)_i = -(\nabla p)_i - \nabla \cdot (\rho \textbf{v}v_i).
\end{align}

Let's expand the RHS in equation (\ref{eq:arf_t_app}):

\begin{align}
    (\nabla p)_i = \frac{\partial p}{\partial x_i}.
\end{align}

\begin{align}\label{eq:arf_t_explicit_app}
\begin{split}
    \nabla \cdot (\rho \textbf{v}v_i) &= \nabla \cdot (\rho v_iv_i\hat{i} + \rho v_iv_j\hat{j} + \rho v_iv_k\hat{k})\\ 
    &= \rho \bigg(\frac{\partial}{\partial x_i}(v_i^2)  + \frac{\partial}{\partial x_j}(v_iv_j) + \frac{\partial}{\partial x_k}(v_iv_k) \bigg) \\ 
    &= \rho\bigg(2v_i\frac{\partial v_i}{\partial x_i} + v_i\frac{\partial v_j}{\partial x_j} \\
    &+ v_j\frac{\partial v_i}{\partial x_j} + v_i\frac{\partial v_k}{\partial x_k} + v_k\frac{\partial v_i}{\partial x_k}\bigg). \\ 
\end{split}
\end{align}

Since our simulations were in 2D, we dropped the last two terms corresponding to a $3^{rd}$ dimension in the above equation. We can rewrite equation (\ref{eq:arf_t_explicit_app}) in 2D more compactly as follows:

\begin{equation}\label{eq:arf_t_compact_app}
    ARF(t)_i = -\frac{\partial}{\partial x_k}\bigg(p\delta_{ik} + \rho v_iv_k \bigg),
\end{equation}

\noindent where $\delta_{ik}$ is the Kronecker delta and repeated indices in $k$ follow Einstein summation. Note that equation (\ref{eq:arf_t_compact_app}) is a function of time. Given the periodicity of the wave, in order to compute the net ARF, we can simply integrate this equation over one full period: 

\begin{align}
\begin{split}
    ARF_i &= \int_{t'}^{t' + \pi} ARF(t)_idt\\
    &= -\int_{t'}^{t'+\pi}\frac{\partial}{\partial x_k}(p\delta_{ik} + \rho v_iv_k)dt \\
    &= -\frac{\partial}{\partial x_k}\int_{t'}^{t'+\pi}(p\delta_{ik} + \rho v_iv_k)dt,
\end{split}
\end{align}

\noindent with $\pi$ denoting one period. Let $\int_{t'}^{t' + \pi}(\cdot)dt = \langle \cdot \rangle$:

\begin{align}\label{eq:arf_net_appendix}
    ARF_i &= -\frac{\partial}{\partial x_k} \langle p\delta_{ik} + \rho v_iv_k\rangle\\ \notag
    &= -\frac{\partial}{\partial x_k}( \langle p\rangle\delta_{ik}  + \langle \rho v_iv_k \rangle),
\end{align}

\noindent where the second term is the Reynolds stress tensor. $\rho$ can be replaced with the unperturbed density, $\rho_0$. Since the unperturbed pressure, $p_0$, is fixed, the pressure term in (\ref{eq:arf_net_appendix}) could be replaced with $\langle p-p_0\rangle $. This term, which we will denote as $\langle P_E\rangle$, is the mean Eulerian excess pressure \cite{Prieur2017ModelingElastography, Lee1993AcousticPressure}. Invoking Taylor's expansion and the first law of thermodynamics, we can write the first term in equation (\ref{eq:arf_net_appendix}) as \cite{Lee1993AcousticPressure}:

\begin{align}\label{eq:EulerianEP_app}
\begin{split}
    \langle p \rangle &= \langle p - p_0\rangle \\ 
    &= \langle P_E\rangle \\ 
    &= -\frac{1}{2}\rho_0 \langle |v|^2 \rangle + \frac{1}{2}\frac{\rho_0}{c^2}\langle \big(\frac{-p}{\rho_0}\big)^2\rangle.
\end{split}
\end{align}

Putting it all together, the $i^{th}$ component of the net ARF is:

\begin{equation}
\begin{split}
    & ARF_i = \\
    &-\frac{\partial}{\partial x_k} \bigg[\bigg(\frac{1}{2\rho_0c^2}\langle p^2 \rangle - \frac{1}{2}\rho_0\langle |v|^2\rangle \bigg)\delta_{ik} + \langle \rho_0 v_iv_k\rangle \bigg].
\end{split}
\end{equation}

Because the simulations in this manuscript were done in Cartesian coordinates in 2D, the explicit form of ARF in $x$ and $y$ was computed as follows:

\begin{align}\label{eq:arf_net_xy_app}
\begin{split}
    ARF_x &= -\frac{\partial}{\partial x} \bigg(\frac{1}{2\rho_0c^2}\langle p^2 \rangle - \frac{1}{2}\rho_0\langle v_x^2 + v_y^2\rangle \bigg) \\
    &- \rho_0 \big\langle  2v_x\frac{\partial v_x}{\partial x} + v_x\frac{\partial v_y}{\partial y} + v_y\frac{\partial v_x}{\partial y} \big\rangle \\
    ARF_y &= -\frac{\partial}{\partial y} \bigg(\frac{1}{2\rho_0c^2}\langle p^2 \rangle - \frac{1}{2}\rho_0\langle v_x^2 + v_y^2\rangle \bigg) \\
    &- \rho_0 \big\langle  2v_y\frac{\partial v_y}{\partial y} + v_y\frac{\partial v_x}{\partial x} + v_x\frac{\partial v_y}{\partial x} \big\rangle
\end{split}
\end{align}

\subsection*{Where Does $2\alpha I / c$ Come From?}

ARF under the purely planar wave assumption assumes the form presented in equation (\ref{eq:ARF_planar}). Under this assumption, pressure takes on the form $p=\rho cv$, with $c$ and $v$ denoting sound and particle velocities, respectively:

\begin{align}\label{eq:2alphaI_initi}
\begin{split}
    ARF^{\text{PP}} &= -\rho \langle \frac{\partial }{\partial x} v_x^2\rangle \\ 
    &= -\rho \frac{\partial}{\partial x}\langle v_x^2 \rangle \\ 
    &= -\rho \frac{\partial}{\partial x} \langle  \frac{p^2}{\rho ^2 c^2} \rangle \text{\ \ since $p=\rho cv$}\\ 
    &= -\frac{\rho}{\rho c}\frac{\partial }{\partial x} \langle \frac{p^2}{\rho c} \rangle \\ 
    &= -\frac{1}{c}\frac{\partial}{\partial x}I_{\text{TA}} \\
    & \text{since temporal average intensity, $I_{\text{TA}}$, is} \\
    & \text{equivalent to $\langle p^2/\rho c\rangle$}.
\end{split}
\end{align}

Note that the gradient of $I_{\text{TA}}$ necessitates an intensity field that is not constant in space. Otherwise the gradient would return zero. Exponential decay of the temporal average intensity is one popular assumption: $I(x)=I_0e^{-2\alpha x}$ with $\alpha$ being the attenuation constant. Substituting for $I_{\text{TA}}$ in equation (\ref{eq:2alphaI_initi}):

\begin{align}\label{eq:2alphaI}
    ARF^{\text{PP}} &= -\frac{1}{c}\frac{\partial }{\partial x}\big(I_0e^{-2\alpha x} \big)\\ \notag
    &= -\frac{1}{c} (-2\alpha I_{\text{TA}})\\ \notag 
    &= \frac{2\alpha I}{c}.
\end{align}

Note that $I$ in equation (\ref{eq:2alphaI}) is the temporal average intensity.

A few words should be said about equation (\ref{eq:2alphaI}) in comparison to equation (\ref{eq:arf_net_xy_app}). First, attenuation is not a necessary condition for momentum transfer. In fact, non-dissipative derivations are common and valid \cite{Prieur2016Simulationk-Wave, Prieur2017ModelingElastography, Sarvazyan2021AcousticApplications}, due to the cumbersome complexities that arise with inclusion of viscosity. As we saw earlier, momentum transfer builds upon conservation of energy. As such, when we remove the viscous forces and arrive at the inviscid formulation in equation (\ref{eq:integral_momentum_inviscid_app}), we still end up computing ARF accurately, without including attenuation. This is due to the fact that in the context of focused ultrasound, normal stresses are dominant in contrast to shear stresses and attenuation due to viscosity can safely be ignored \cite{Prieur2017ModelingElastography}. Second, although the attenuation term in equation (\ref{eq:2alphaI}) makes it seem necessary for the existence of ARF, it is only necessary if the purely planar wave assumption is considered. Under such extreme simplifying assumptions, where planar waves are equally present everywhere in space, not including attenuation would simply return a zero gradient for intensity. Of course one could first run a simulation to steady state and instead of using a simple exponentially attenuated formula for intensity,  directly take the gradient of the accurately simulated temporal average intensity, $-\frac{1}{c}\frac{\partial}{\partial x}I_{\text{TA}}$. This way, unlike the exponentially attenuated beam assumption, the maximum beam intensity would accurately fall on the focus. However, as we saw in the simulations for the purely planar wave assumption, this simple gradient would result in a categorically false ARF field with part of the force components pointing towards the transducer surface and part pointing away from the transducer. Third, we could fine tune the $\alpha$ parameter so that the numerical value of ARF matches that computed with equation (\ref{eq:arf_net_xy_app}). However, even then, equation (\ref{eq:2alphaI}) fails to provide any notion of direction for ARF. In simple, single transducer scenarios, one could augment the fine tuning idea with the intuitive understanding that any value computed for ARF should point away from the transducer surface. That way, a notion of direction could be ascribed to the numerical value computed via equation (\ref{eq:2alphaI}). However, in more complex situations such as the orthogonal setup, prior knowledge of the directionality of ARF components is rather difficult if not impossible. To that end, we highly encourage the use of equation (\ref{eq:arf_net_xy_app}) instead of equation (\ref{eq:2alphaI}). Although equation (\ref{eq:arf_net_xy_app}) is written out explicitly for 2D, its extension to 3D is trivial: 

\begin{equation}\label{ARF_3D}
\begin{split}
    ARF_i &= -\frac{\partial}{\partial x_i} \bigg(\frac{1}{2\rho_0c^2}\langle p^2 \rangle - \frac{1}{2}\rho_0\langle v_i^2 + v_j^2 + v_k^2 \rangle \bigg) \\
    &- \rho_0 \big\langle  2v_i\frac{\partial v_i}{\partial x_i} + v_i\frac{\partial v_j}{\partial x_j} + v_j\frac{\partial v_i}{\partial x_j} + v_k\frac{\partial v_i}{\partial x_k} + v_i\frac{\partial v_k}{\partial x_k} \big\rangle.
\end{split}
\end{equation}

\section*{Code Availability}
All simulation files and analysis scripts are available at: \url{https://github.com/kbp-lab/Crossbeam}.

\bibliographystyle{ieeetr}

\begin{thebibliography}{99}

\bibitem{FRY1958IntenseSystem.}
W.~J. FRY, ``{Intense ultrasound in investigations of the central nervous system.},'' {\em Advances in biological and medical physics}, vol.~6, 1958.

\bibitem{Yang2018NeuromodulationDetection}
P.~F. Yang, M.~A. Phipps, A.~T. Newton, V.~Chaplin, J.~C. Gore, C.~F. Caskey, and L.~M. Chen, ``{Neuromodulation of sensory networks in monkey brain by focused ultrasound with MRI guidance and detection},'' {\em Scientific Reports}, vol.~8, no.~1, 2018.

\bibitem{Folloni2019ManipulationStimulation}
D.~Folloni, L.~Verhagen, R.~B. Mars, E.~Fouragnan, C.~Constans, J.~F. Aubry, M.~F. Rushworth, and J.~Sallet, ``{Manipulation of Subcortical and Deep Cortical Activity in the Primate Brain Using Transcranial Focused Ultrasound Stimulation},'' {\em Neuron}, vol.~101, no.~6, 2019.

\bibitem{Kim2021TranscranialSheep}
H.~C. Kim, W.~Lee, J.~Kunes, K.~Yoon, J.~E. Lee, L.~Foley, K.~Kowsari, and S.~S. Yoo, ``{Transcranial focused ultrasound modulates cortical and thalamic motor activity in awake sheep},'' {\em Scientific Reports}, vol.~11, no.~1, 2021.

\bibitem{Legon2018NeuromodulationThalamus}
W.~Legon, L.~Ai, P.~Bansal, and J.~K. Mueller, ``{Neuromodulation with single-element transcranial focused ultrasound in human thalamus},'' {\em Human Brain Mapping}, vol.~39, no.~5, 2018.

\bibitem{Mohammadjavadi2022TranscranialMR-ARFI}
M.~Mohammadjavadi, R.~T. Ash, N.~Li, P.~Gaur, J.~Kubanek, Y.~Saenz, G.~H. Glover, G.~R. Popelka, A.~M. Norcia, and K.~B. Pauly, ``{Transcranial ultrasound neuromodulation of the thalamic visual pathway in a large animal model and the dose-response relationship with MR-ARFI},'' {\em Scientific Reports}, vol.~12, no.~1, 2022.

\bibitem{Zhang2021TranscranialAnimals}
T.~Zhang, N.~Pan, Y.~Wang, C.~Liu, and S.~Hu, ``{Transcranial Focused Ultrasound Neuromodulation: A Review of the Excitatory and Inhibitory Effects on Brain Activity in Human and Animals},'' {\em Frontiers in Human Neuroscience}, vol.~15, 9 2021.

\bibitem{Sanguinetti2020TranscranialHumans}
J.~L. Sanguinetti, S.~Hameroff, E.~E. Smith, T.~Sato, C.~M. Daft, W.~J. Tyler, and J.~J. Allen, ``{Transcranial Focused Ultrasound to the Right Prefrontal Cortex Improves Mood and Alters Functional Connectivity in Humans},'' {\em Frontiers in human neuroscience}, vol.~14, 2 2020.

\bibitem{MartinUltrasoundCircuits}
E.~Martin, M.~Roberts, I.~F. Grigoras, O.~Wright, T.~Nandi, S.~W. Rieger, J.~Campbell, T.~Den Boer, B.~T. Cox, C.~J. Stagg, and B.~E. Treeby, ``{Ultrasound system for precise neuromodulation of human deep brain circuits},'' {\em bioRxiv}, 2024.

\bibitem{MacKinnon2022Sensitivity}
N.~MacKinnon, J.~G. Korvink, and M.~Jouda, ``{ Microengineering Improves MR Sensitivity },'' in {\em Magnetic Resonance Microscopy}, 2022.


\bibitem{chen2004comparison}
S. Chen, M. Fatemi, R. Kinnick, and J. F. Greenleaf, "Comparison of stress field forming methods for vibro-acoustography," \textit{IEEE Transactions on Ultrasonics, Ferroelectrics, and Frequency Control}, vol. 51, pp. 313--321, 2004. DOI: 10.1109/TUFFC.2004.1320787.


\bibitem{Kim2018ImprovedTransducers}
S.~Kim, H.~Kim, C.~Shim, and H.~J. Lee, ``{Improved Target Specificity of Transcranial Focused Ultrasound Stimulation (TFUS) using Double-Crossed Ultrasound Transducers},'' in {\em Proceedings of the Annual International Conference of the IEEE Engineering in Medicine and Biology Society, EMBS}, vol.~2018-July, 2018.

\bibitem{Yang2020DevelopmentStudy}
Y.~Yang, C.~Wang, Y.~Li, J.~Huang, F.~Cai, Y.~Xiao, T.~Ma, and H.~Zheng, ``{Development of Scalable 2D Plane Array for Transcranial Ultrasonic Neuromodulation on Non-Human Primates: An Ex Vivo Study},'' {\em IEEE Transactions on Neural Systems and Rehabilitation Engineering}, vol.~28, no.~2, 2020.

\bibitem{Kim2021TranscranialResolution}
S.~Kim, Y.~Jo, G.~Kook, C.~Pasquinelli, H.~Kim, K.~Kim, H.~S. Hoe, Y.~Choe, H.~Rhim, A.~Thielscher, J.~Kim, and H.~J. Lee, ``{Transcranial focused ultrasound stimulation with high spatial resolution},'' {\em Brain Stimulation}, vol.~14, no.~2, 2021.

\bibitem{Menz2019RadiationRetina}
M.~D. Menz, P.~Ye, K.~Firouzi, A.~Nikoozadeh, K.~B. Pauly, P.~Khuri-Yakub, and S.~A. Baccus, ``{Radiation force as a physical mechanism for ultrasonic neurostimulation of the ex vivo retina},'' {\em Journal of Neuroscience}, vol.~39, no.~32, 2019.

\bibitem{Kim2022PatternedNeuromodulation}
Y.~H. Kim, K.~C. Kang, J.~N. Kim, C.~N. Pai, Y.~Zhang, P.~Ghanouni, K.~K. Park, K.~Firouzi, and B.~T. Khuri-Yakub, ``{Patterned Interference Radiation Force for Transcranial Neuromodulation},'' {\em Ultrasound in Medicine and Biology}, vol.~48, no.~3, 2022.

\bibitem{Wright2017UnmyelinatedUltrasound}
C.~J. Wright, S.~R. Haqshenas, J.~Rothwell, and N.~Saffari, ``{Unmyelinated Peripheral Nerves Can Be Stimulated in¬†Vitro Using Pulsed Ultrasound},'' {\em Ultrasound in Medicine and Biology}, vol.~43, no.~10, 2017.

\bibitem{Kubanek2018UltrasoundSystem}
J.~Kubanek, P.~Shukla, A.~Das, S.~A. Baccus, and M.~B. Goodman, ``{Ultrasound elicits behavioral responses through mechanical effects on neurons and ion channels in a simple nervous system},'' {\em Journal of Neuroscience}, vol.~38, no.~12, 2018.

\bibitem{Blackmore2019UltrasoundSafety}
J.~Blackmore, S.~Shrivastava, J.~Sallet, C.~R. Butler, and R.~O. Cleveland, ``{Ultrasound Neuromodulation: A Review of Results, Mechanisms and Safety},'' 2019.

\bibitem{Prieto2013DynamicForce}
M.~L. Prieto, Ö. Oralkan, B.~T. Khuri-Yakub, and M.~C. Maduke, ``{Dynamic Response of Model Lipid Membranes to Ultrasonic Radiation Force},'' {\em PLoS ONE}, vol.~8, no.~10, 2013.


\bibitem{Kamimura2020UltrasoundAssessment}
H.~A. Kamimura, A.~Conti, N.~Toschi, and E.~E. Konofagou, ``{Ultrasound neuromodulation: Mechanisms and the potential of multimodal stimulation for neuronal function assessment},'' 2020.

\bibitem{Palmeri2011AcousticMethods}
M.~L. Palmeri and K.~R. Nightingale, ``{Acoustic radiation force-based elasticity imaging methods},'' 2011.

\bibitem{Hsu2005AcousticProbe}
S.~J. Hsu, B.~J. Fahey, D.~M. Dumont, and G.~E. Trahey, ``{Acoustic radiation force impulse imaging with an intra-cardiac probe},'' in {\em Medical Imaging 2005: Ultrasonic Imaging and Signal Processing}, vol.~5750, 2005.

\bibitem{Mcaleavey2009ValidationHydrogels}
S.~Mcaleavey, E.~Collins, J.~Kelly, E.~Elegbe, and M.~Menon, ``{Validation of SMURF Estimation of Shear Modulus in Hydrogels},'' {\em Ultrasonic Imaging}, vol.~31, no.~1, 2009.

\bibitem{Haar2010UltrasoundSafety}
G.~T. Haar, ``{Ultrasound bioeffects and safety},'' {\em Proceedings of the Institution of Mechanical Engineers, Part H: Journal of Engineering in Medicine}, vol.~224, no.~2, 2010.

\bibitem{Nightingale2003Shear-waveResults}
K.~Nightingale, S.~McAleavey, and G.~Trahey, ``{Shear-wave generation using acoustic radiation force: In vivo and ex vivo results},'' {\em Ultrasound in Medicine and Biology}, vol.~29, no.~12, 2003.

\bibitem{Nightingale2006AnalysisForce}
K.~Nightingale, M.~Palmeri, and G.~Trahey, ``{Analysis of contrast in images generated with transient acoustic radiation force},'' {\em Ultrasound in Medicine and Biology}, vol.~32, no.~1, 2006.

\bibitem{Doherty2013AcousticUltrasound}
J.~R. Doherty, G.~E. Trahey, K.~R. Nightingale, and M.~L. Palmeri, ``{Acoustic radiation force elasticity imaging in diagnostic ultrasound},'' {\em IEEE Transactions on Ultrasonics, Ferroelectrics, and Frequency Control}, vol.~60, no.~4, 2013.

\bibitem{Zhang2018FromForces}
L.~Zhang, ``{From acoustic radiation pressure to three-dimensional acoustic radiation forces},'' {\em The Journal of the Acoustical Society of America}, vol.~144, no.~1, 2018.

\bibitem{Sarvazyan2021AcousticApplications}
A.~P. Sarvazyan, O.~V. Rudenko, and M.~Fatemi, ``{Acoustic Radiation Force: A Review of Four Mechanisms for Biomedical Applications},'' 2021.

\bibitem{Treeby2010K-Wave:Fields}
B.~E. Treeby and B.~T. Cox, ``{k-Wave: MATLAB toolbox for the simulation and reconstruction of photoacoustic wave fields},'' {\em Journal of Biomedical Optics}, vol.~15, no.~2, 2010.

\bibitem{TheMathWorksInc2019MATLABR2019b}
{The MathWorks Inc}, ``{MATLAB version: 9.7.0 (R2019b)},'' 2019.

\bibitem{Fink1994PhaseMirrors}
M.~Fink and C.~Dorme, ``Phase aberration correction with ultrasonic time
  reversal mirrors,'' \emph{Proceedings of the IEEE Ultrasonics Symposium},
  vol.~3, pp. 1629--1638, 1994.
  
\bibitem{Prieur2016Simulationk-Wave}
F.~Prieur and S.~Catheline, ``Simulation of shear wave elastography imaging using the toolbox k-Wave,'' in {\em Proceedings of Meetings on Acoustics}, vol.~29, 2016.

\bibitem{MarkDrela2005FluidMechanics}
{Mark Drela}, ``{Fluid Mechanics},'' 2005.

\bibitem{JohnD.Anderson2001FundamentalsAerodynamics}
{John D. Anderson}, {\em {Fundamentals of Aerodynamics}}.
\newblock 2001.

\bibitem{Prieur2017ModelingElastography}
F.~Prieur and O.~A. Sapozhnikov, ``{Modeling of the acoustic radiation force in elastography},'' {\em The Journal of the Acoustical Society of America}, vol.~142, no.~2, 2017.

\bibitem{Lee1993AcousticPressure}
C.~P. Lee and T.~G. Wang, ``{Acoustic radiation pressure},'' {\em Journal of the Acoustical Society of America}, vol.~94, no.~2, 1993.

\bibitem{Batchelor2000AnDynamics}
G.~K. Batchelor, {\em {An Introduction to Fluid Dynamics}}.
\newblock 2000.

\bibitem{Treeby2014ModellingToolbox}
B.~E. Treeby, J.~Jaros, D.~Rohrbach, and B.~T. Cox, ``{Modelling elastic wave propagation using the k-Wave MATLAB Toolbox},'' in {\em IEEE International Ultrasonics Symposium, IUS}, 2014.

\bibitem{Carstensen2016BiologicalDescriptors}
E.~L. Carstensen, K.~J. Parker, D.~Dalecki, and D.~C. Hocking, ``{Biological Effects of Low-Frequency Shear Strain: Physical Descriptors},'' 2016.

\bibitem{McDannold2008MagneticImaging}
N.~McDannold and S.~E. Maier, ``{Magnetic resonance acoustic radiation force imaging},'' {\em Medical Physics}, vol.~35, no.~8, 2008.

\bibitem{Duryea2011InHistotripsy}
A.~P. Duryea, A.~D. Maxwell, W.~W. Roberts, Z.~Xu, T.~L. Hall, and C.~A. Cain, ``{In vitro comminution of model renal calculi using histotripsy},'' {\em IEEE Transactions on Ultrasonics, Ferroelectrics, and Frequency Control}, vol.~58, no.~5, 2011.

\bibitem{Roberts2005FocusedDirections}
W.~W. Roberts, ``{Focused ultrasound ablation of renal and prostate cancer: Current technology and future directions},'' in {\em Urologic Oncology: Seminars and Original Investigations}, vol.~23, 2005.

\bibitem{Roberts2014DevelopmentDirections}
W.~W. Roberts, ``{Development and translation of histotripsy: Current status and future directions},'' 2014.

\bibitem{Xu2004ControlledErosion}
Z.~Xu, A.~Ludomirsky, L.~Y. Eun, T.~L. Hall, B.~C. Tran, J.~B. Fowlkes, and C.~A. Cain, ``{Controlled ultrasound tissue erosion},'' {\em IEEE Transactions on Ultrasonics, Ferroelectrics, and Frequency Control}, vol.~51, no.~6, 2004.

\bibitem{Riis2022Multifrequency-basedVolume}
T.~Riis and J.~Kubanek, ``{Multifrequency-based sharpening of focal volume},'' {\em Scientific Reports}, vol.~12, no.~1, 2022.

\bibitem{King2013EffectiveNeurostimulation}
R.~L. King, J.~R. Brown, W.~T. Newsome, and K.~B. Pauly, ``{Effective parameters for ultrasound-induced in vivo neurostimulation},'' {\em Ultrasound in medicine {\&} biology}, vol.~39, no.~2, pp.~312--331, 2013.

\bibitem{Tufail2010TranscranialCircuits}
Y.~Tufail, A.~Matyushov, N.~Baldwin, M.~L. Tauchmann, J.~Georges, A.~Yoshihiro, S.~I. Tillery, and W.~J. Tyler, ``{Transcranial pulsed ultrasound stimulates intact brain circuits},'' {\em Neuron}, vol.~66, pp.~681--694, 6 2010.

\bibitem{Legon2014TranscranialHumans}
W.~Legon, T.~F. Sato, A.~Opitz, J.~Mueller, A.~Barbour, A.~Williams, and W.~J. Tyler, ``{Transcranial focused ultrasound modulates the activity of primary somatosensory cortex in humans},'' {\em Nature Neuroscience 2013 17:2}, vol.~17, pp.~322--329, 1 2014.

\bibitem{Lee2015Image-GuidedCortex}
W.~Lee, H.~Kim, Y.~Jung, I.~U. Song, Y.~A. Chung, and S.~S. Yoo, ``{Image-Guided Transcranial Focused Ultrasound Stimulates Human Primary Somatosensory Cortex},'' {\em Scientific Reports 2015 5:1}, vol.~5, pp.~1--10, 3 2015.

\bibitem{Schey2004DivCalculus}
H.~M. Schey, {\em {Div Grad Curl and All That an Informal Text on Vector Calculus}}.
\newblock 2004.

\end{thebibliography}

\end{document}